\documentclass[prl,twocolumn,superscriptaddress]{revtex4-2}
\usepackage{amsmath,amssymb,bm}
\usepackage{hyperref}
\usepackage{graphicx}
\usepackage{epstopdf}
\usepackage{latexsym}
\usepackage{subfigure}
\usepackage[usenames, dvipsnames]{color}
\usepackage[usenames, dvipsnames]{xcolor}
\usepackage{natbib}
\usepackage{braket}
\usepackage{float}
\usepackage[normalem]{ulem}
\usepackage{comment}
\usepackage{mathtools}
\usepackage{array}
\usepackage{tabu}
\usepackage{multirow}
\usepackage{chemformula}
\usepackage{svg}
\usepackage{tabularx,ragged2e}
\usepackage{dcolumn}

\newcommand\redsout{\bgroup\markoverwith{\textcolor{red}{\rule[0.5ex]{2pt}{0.4pt}}}\ULon}
\newcommand\bluesout{\bgroup\markoverwith{\textcolor{blue}{\rule[0.5ex]{2pt}{0.4pt}}}\ULon}

\newcommand{\prlsec}[1]{{{\it #1:--}}}

\newcommand{\SPhide}[1]{{}}

\newcommand{\NDhide}[1]{{}}

\begin{document}
\title{Deconfined pseudocriticality in a model spin-1 quantum antiferromagnet}
\author{Vikas Vijigiri}
\affiliation{Department of Physics, Indian Institute of
Technology Bombay, Powai, Mumbai, MH 400076, India}
\author{Sumiran Pujari}
\email{sumiran.pujari@iitb.ac.in}
\affiliation{Department of Physics, Indian Institute of
Technology Bombay, Powai, Mumbai, MH 400076, India}
\author{Nisheeta Desai}
\affiliation{Department of Theoretical Physics, Tata Institute
of Fundamental Research, Colaba, Mumbai, MH 400005, India}

\begin{abstract}
Berry phase interference arguments that underlie the theory of 
deconfined quantum criticality (DQC) for $S=1/2$ antiferromagnets have 
also been invoked to allow for 
continuous transitions in $S=1$ magnets
including a N\'eel to (columnar) valence bond solid (cVBS)
transition. 
We provide a microscopic model realization of this transition
on the square lattice consisting of 
Heisenberg exchange ($J_H$) and biquadratic exchange ($J_B$) that
favor a N\'eel phase, and
 a designed $Q$-term ($Q_B$) interaction which favors a cVBS
 through large-scale quantum Monte Carlo (QMC) simulations. 
 For $J_H=0$, this model is equivalent to the $SU(3)$ 
 $JQ$ model 
 with a N\'eel-cVBS transition that has been argued to be 
 DQC through QMC.
 Upon turning on $J_H$ which brings down the symmetry to $SU(2)$, we find 
 multiple signatures -- a single critical point,
high quality collapse of correlation ratios and order parameters,
``$U(1)$-symmetric" cVBS histograms and
lack of double-peak in order parameter histograms 
for largest sizes studied
near the critical point --
that are highly suggestive of a continuous transition scenario.
However, Binder analysis finds negative dips that grow sub-extensively
that we interpret as these
transitions rather being pseudocritical.
This along with recent results on spin-$\frac{1}{2}$ models
suggests that deconfined pseudocriticality is 
the more generic scenario.

\end{abstract}
\maketitle

The theory of deconfined quantum criticality 
(DQC)~\cite{Senthil_etal_Science_2004,Senthil_etal_PRB_2004,levin_senthil_prb2004,
Sandvik_Scalapino_etal_PRL_2002} 
has been of great interest 
as it lies beyond the Landau-Ginzburg-Wilson-Fisher paradigm.
It posits a continuous transition between two symmetry unrelated phases
-- N\'eel and valence bond solid (VBS) -- for spin-$\frac{1}{2}$
moments in $2+1d$. 
DQC is described as a gauge theory of fractionalized 
spinon degrees of freedom that deconfine only at the critical point.
Their Higgs condensation 
leads to antiferromagnetic N\'eel order,
while on the other side the confinement of the 
associated $U(1)$ gauge field~\cite{Haldane_prl1988,Read_Sachdev_prl1989,Read_Sachdev_prb1990,Read_Sachdev_nucphysB} 
leads to the VBS.
It has undergone a 
great deal of scrutiny~\cite{Kaul_Melko_Sandvik_2013} 
in various spin-$1/2$ models~\cite{Sandvik_prl2007, Melko_Kaul_prl2008, Sandvik_prl2010, Banerjee_Damle_Alet_prb2010, Banerjee_Damle_Alet_prb2011, Sandvik_prb2012, Kaul_prb2014, Pujari_Damle_Alet_prl2013, Pujari_Alet_Damle_prb2015,Ma_etal_prl2019,Zhao_Takahashi_Sandvik_prl2020,Sandvik_Zhao_2020} and their SU($N$) generalizations~\cite{Beach_etal_prb2009, Lou_Sandvik_Kawashima_PRB_2009, Kaul_prb2012, Kaul_Sandvik_prl2012, Block_Melko_Kaul_prl2013, Harada_etal_prb2013}.
Evidence for many features of DQC have been numerically
seen in these studies,
including classical loop models and dimer models in 3$d$~\cite{charrier_prl2008,Powell_prl2008,Powell_prb2009,Chen_prb2009},
certain $1+1d$ spin-$\frac{1}{2}$ extensions~\cite{Roberts_Jiang_Motrunich_prb2019,Huang_etal_prb2019,
Zhang_Levin_PRL2023},
and fermionic models~\cite{DaLiao_etal_prbAug2022,DaLiao_etal_prbOct2022,Sato_etal_prl2017,Li_natcomm2017,Li_arxiv2019}.However scaling violations 
have also been seen~\cite{Kuklov_2006, Jiang_JStatMech2008,  Kaul_prb2011,  Kun_prl2013, Bartosch_PRB_2013, Shao_Guo_Sandvik_Science2016, nahum_etal_prx2015} 
that are still under debate. 
Not much is known though for $S>\frac{1}{2}$.

Our focus will be on spin-$1$ here.
There are only a handful of works discussing possible DQC
and none which have shown DQC behavior 
in microscopic model realizations.
Previously, Ref.~\cite{Grover_Senthil_PRL_2007}
argued for a possible DQC from a spin-nematic state
to a VBS state
based on field-theoretic arguments.
Ref.~\cite{Wang_Kivelson_Lee_NatPhys_2015} 
similarly conjectured a possible DQC from
N\'eel to a bond-nematic or Haldane-nematic state
which has been numerically investigated in
Refs.~\cite{Jiang_etal_PRB_2009,Niesen_Corboz_PRB_2017, Niesen_Corboz_SciPost_2017,
Hu_etal_PRB_2019}.
It further conjectured possible DQC 
from N\'eel to cVBS as a ``doubled" 
spin-$\frac{1}{2}$ DQC theory~\cite{doubled_theory_footnote}.
Briefly, the theory is formulated by taking 
two copies of an $SO(5)$ field theory with
$k=1$ Wess-Zumino-Witten term
for a combined 5-component order parameter field made
from the N\'eel and columnar-VBS 
fields~\cite{Tanaka_Hu_prl2005,Senthil_Fisher_prb2006}
that has been used to describe $S=\frac{1}{2}$ DQC. 
Upon
invoking a strong ferromagnetic coupling between the two
copies to be consistent with spin-$1$ at low energies, 
it reduces to a ``single" $SO(5)$ field theory
now with a doubled Wess-Zumino-Witten term. It is
the presence of this topological term which can allow for a
DQC betwen N\'eel to cVBS (for details, see the
supplementary of Ref.~\cite{Wang_Kivelson_Lee_NatPhys_2015}).

We computationally investigate this latter scenario 
based on the following heuristic:
$JQ$ models~\cite{Sandvik_prl2007} 
have been crucial in probing DQC physics 
in large-scale simulations. The basic units are 
$SU(2)$ singlet projectors $P^2_{ij} = \left(\frac{1}{4} -
\mathbf{s}_i \cdot \mathbf{s}_j \right)$ for $S=\frac{1}{2}$
on bond $\langle ij \rangle$.
One can extend~\cite{sublattice_footnote} them 
to $SU(N)$ as 
$P^N_{ij} = \sum^N_{\alpha,\beta=1} |\alpha_i;\alpha_j\rangle\langle\beta_i;\beta_j|$.
The $U(1)$-gauge fluctuations get suppressed as $N$ increases. 
This gives closer match between perturbation theory 
and numerical estimates 
of critical exponents in $N>2$ microscopic models~\cite{Fig5_footnote}. 
The $SU(3)$ $JQ$ model 
\begin{equation}
H_{SU(3)} = -J_B \sum_{\langle ij \rangle} P^3_{ij}
-
Q_B\sum_{\langle ijkl \rangle} \left( P^3_{ij} P^3_{kl} + P^3_{il} P^3_{jk} \right)
\label{eq:SU3Qterm}
\end{equation}
can also be
recasted as a spin-$1$ model 
since $P_{ij} = \frac{1}{3} \{(\mathbf{S}_i \cdot \mathbf{S}_j)^2 - \mathbb{I}\}$
biquadratic exchange for $S=1$~\cite{Beach_etal_prb2009}. 
$\langle ijkl \rangle$ indexes elementary plaquettes of
the square lattice with a clockwise indexing of the sites $i,j,k,l$.
The first term favors N\'eel order. The $Q$-term favors a cVBS of
spin-$1$ valence bonds (also $SU(3)$ singlets
for Eq.~\ref{eq:SU3Qterm}).
DQC behavior
between N\'eel and VBS has been observed in the $SU(3)$ $JQ$ model
up to system sizes $L=48$~\cite{Lou_Sandvik_Kawashima_PRB_2009}.

\begin{figure}[t]
    \centering
    \includegraphics[width=0.99\linewidth]{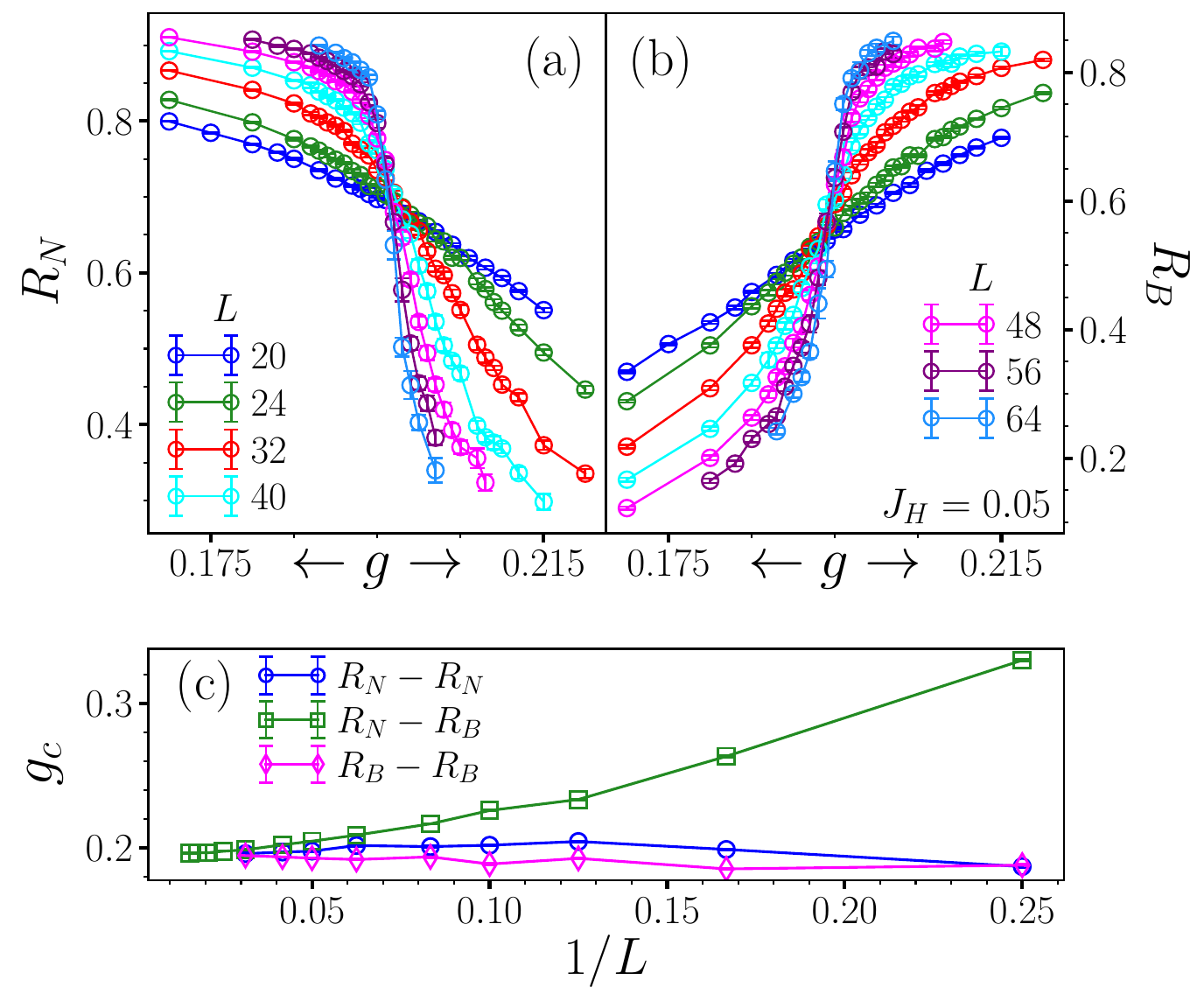}
    \caption{\label{fig:Ratios_gc}
    Correlation ratios of (a) N\'eel and (b) cVBS 
    order parameters ($R_N$ and $R_V$~\cite{corr_ratio_footnote}) 
    $g=Q_{B}/J_{B}$, $J_H=0.05$, $\beta=\frac{L}{4}$. 
    (c) shows $g$ where the $R_N$ values cross for $L$ and $L/2$
    versus $1/L$, similarly for $R_V$, and where $R_N$ and $R_V$ values
    cross for each $L$.
    The three estimates of $g_c(L)$ converge to $g_c \sim 0.195$ implying a direct transition without an intermediate phase.
    }
\end{figure}

We now
add $S=1$ Heisenberg exchange favoring N\'eel order 
as an $SU(2)$-symmetric perturbation 
to study the same transition, 
i.e.
\begin{equation}
H_{SU(2)} = H_{SU(3)} + J_{H} \sum_{\langle ij \rangle} 
\{\mathbf{S}_i \cdot \mathbf{S}_j - \mathbb{I}\}
\label{eq:SU2Ham}
\end{equation}
by varying 
$g \equiv \frac{Q_B}{J_B}$ for different
$J_H$ using  
quantum Monte Carlo (QMC) methods~\cite{Desai_Kaul_prl2019,Sandvik_1991,Sandvik_1992,Sandvik_2010_review}.
We note for later discussion that 
our QMC method work with 
a (``doubled") Hilbert space of two ``split $S=\frac{1}{2}$"s per site
and a symmetrization step~\cite{Kawashima_1995,Todo_2001} to restrict
to the physical $S=1$ subspace quite in analogy with doubled $S=\frac{1}{2}$
DQC theory. 
An earlier work~\cite{Wildeboer_etal_prb2020}
had studied the $J_{B}=0$ case and found 
strong first-order behavior. 
We will thus focus on the vicinity of $H_{SU(3)}$,
i.e. Eq.~\ref{eq:SU2Ham} 
where $i$ refers to sites, $\langle ij\rangle$ to nearest
neighbor bonds, and $\mathbf{S}_i$ to spin-$1$ operators.
We work in the units where $J_B=1.0$. 
The following scenarios 
may be expected: (1) $SU(3)$ criticality becomes first-order right upon turning 
on $J_H$ and we see some cross-over physics for small values of the 
perturbation, (2) there is a regime of $J_H$ for which the doubled 
$S=1/2$ DQC scenario obtains, or (3) there is weakly first-order or 
pseudocritical behaviour in this
regime. For the second scenario, we expect to see stable exponents that are 
either $SU(3)$ exponents or a new set of $SU(2)$ exponents.

We probe the system by measuring intensive order parameters for the 
two phases. For the N\'eel phase, the 
staggered magnetization order parameter is
$O_N \equiv \langle m^2 \rangle$, where 
$m = \frac{1}{N} \sum_{\mathbf r} e^{i (\pi,\pi).\mathbf{r}} S^z_{\mathbf{r}}$. 
For the columnar VBS phase, the cVBS order parameter is
$O_V \equiv \langle \left( \phi^2_x + \phi^2_y \right) \rangle$, 
where $\phi_\mu = \frac{1}{N} \sum_{\mathbf r} 
e^{i \pi \,\mathbf{e}_{\mu} \cdot \mathbf{r}} 
\{ (\mathbf{S}_{\mathbf r} \cdot \mathbf{S}_{\mathbf {r + e_\mu}})^2
- \mathbb{I} \}/3$~\cite{cvbs_footnote}.
The inverse temperature 
$\beta$ is set equal to $L/4$ to study ground state 
properties~\cite{beta_convergence_footnote}.
Fig~\ref{fig:Ratios_gc} shows correlation ratios ($R$) of these order
parameters~\cite{corr_ratio_footnote} 
versus $g=\frac{Q_B}{J_B}$
for a representative value of $J_H=0.05$.
We see a clear crossing at the transition 
in Fig.~\ref{fig:Ratios_gc}a,b. The crossing points of both the N\'eel and VBS 
ratios converge 
versus $1/L$ (Fig.~\ref{fig:Ratios_gc}c)
implying a direct  transition between the two phases with no 
intermediate phase. 

\begin{figure}[t]
 \includegraphics[width=0.95\linewidth]{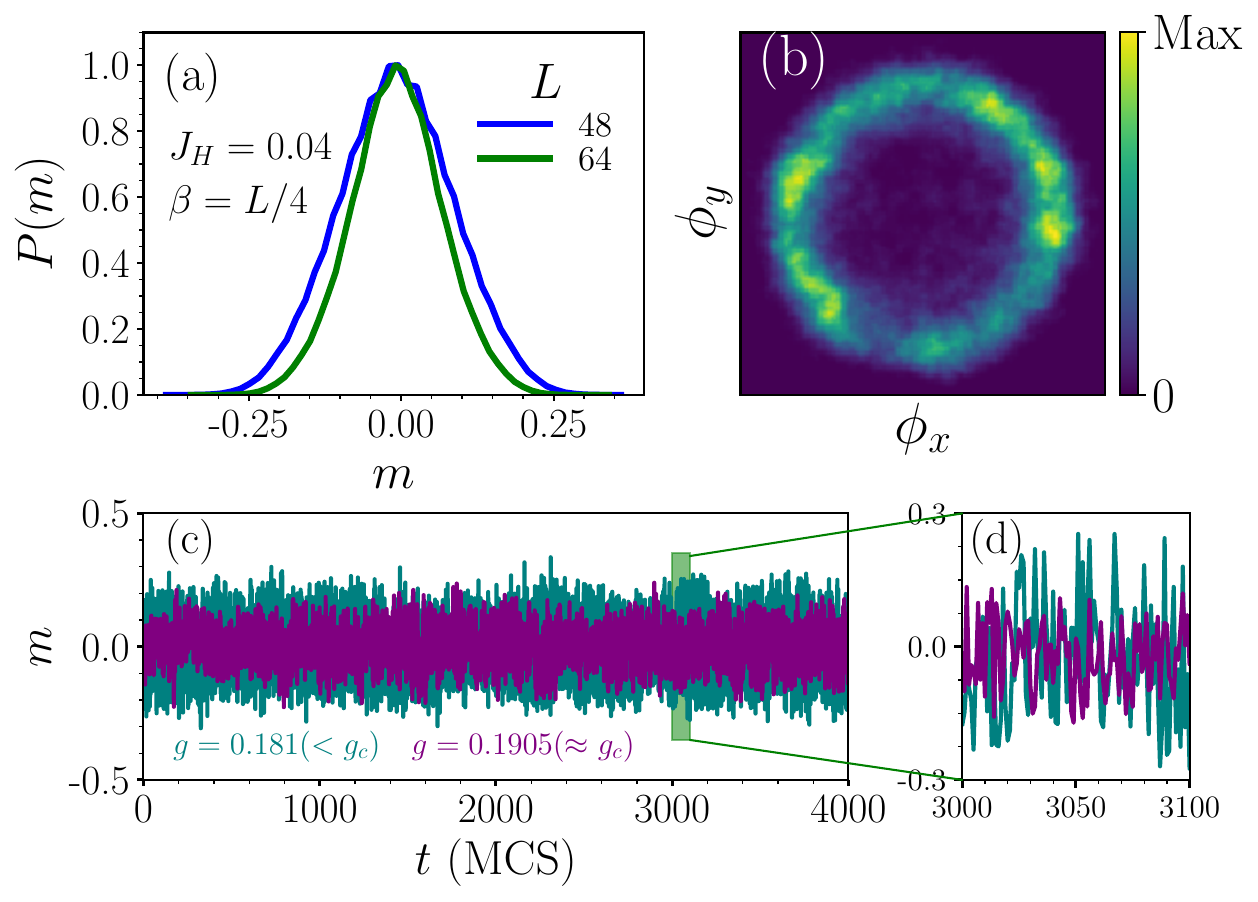}
 \caption{\label{fig:histograms}(a) Histogram of 
 staggered magnetization $m$ (along
 $z$-direction) for $J_H=0.04$ near
 $g_c \sim 0.19$. (b) Heat Map of the 
 cVBS order parameters ($\phi_x,\phi_y$) shows
 ``$U(1)$-symmetry" characteristic of 
 DQC. (c) The time series plot for $m$ 
 does not show any telegraphic switching behaviour 
 ruling out a strong first-order
 transition. $L=64$ in (b),(c).
 }
\end{figure}

\begin{figure}[b]
\includegraphics[width=0.98\linewidth]{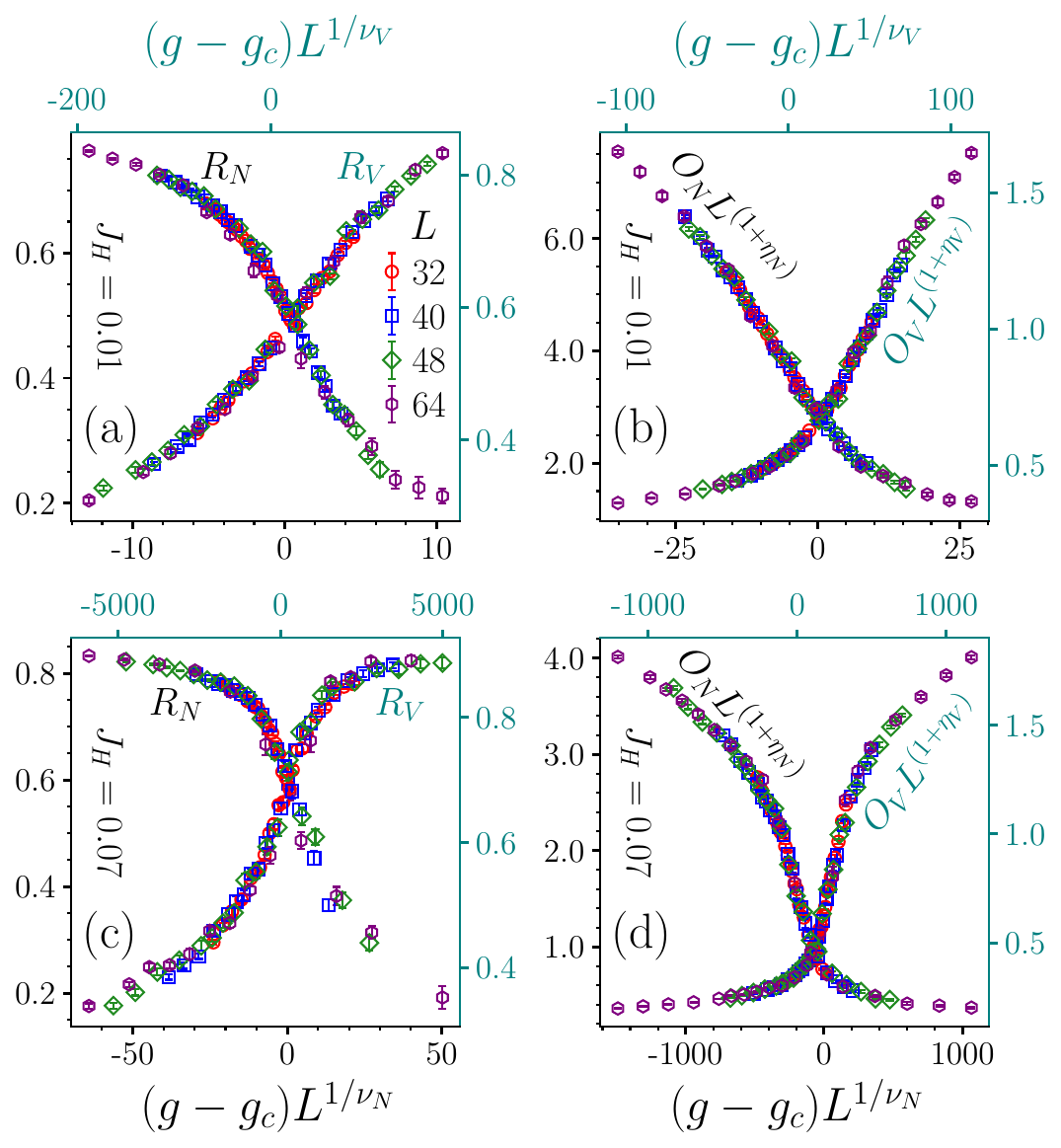}
 \caption{\label{fig:scalingcollapse} 
 Scaling collapse of correlation ratios
 ($R_N$, $R_V$) and order parameters ($O_N$, $O_V$)
 for $J_H=0.01$ and $0.07$ with best estimates 
 of the critical exponents (Table~\ref{tab:exponents}).
 ($\beta=\frac{L}{4}$). Note the high quality of
 collapse. Also associated $\chi^2$ values 
 in Table~\ref{tab:exponents}. This highly suggests 
 a continuous transition within the doubled-DQC framework.
 }
\end{figure}

We now look at order parameter histograms to probe 
the nature of this direct transition. 
No signature of two-peak behavior is seen
near the transition  as shown for staggered magnetization 
in Fig~\ref{fig:histograms}a. 
Telegraphic switching between the two 
order parameters is also absent near 
the transition (Fig~\ref{fig:histograms}c). This
rules out the first scenario of a strongly first-order transition.
Furthermore, ``$U(1)$-symmetric"
$(\phi_x,\phi_y)$-histograms are also seen near the transition
for the largest system size studied
(Fig~\ref{fig:histograms}b). In $S=\frac{1}{2}$ studies,
this has been considered a key evidence of DQC.
This is associated with the dangerous irrelevancy of the 
the operators in the DQC theory that capture the 
dominant quantum fluctuations out of the N\'eel phase.
For $S=1$, the appropriate operators
are those of the doubled-DQC theory~\cite{monopole_footnote}.
We therefore perform scaling collapses~\cite{Harada_PRE2011} 
of the order parameters and
the correlation ratios as shown in Fig.~\ref{fig:scalingcollapse}.
The scaling collapses are of high quality for all $J_H$
with the $\chi^2$ per degree of freedom being close to 1
throughout.
Table~\ref{tab:exponents} lists the exponents extracted from 
this finite-size scaling analysis.
The exponents are stable to various protocols involving 
the range of the tuning parameter $g$ and system sizes 
used for the collapses. 

\begin{table*}
\caption{\label{tab:exponents}
Critical exponents ($\eta_V,\eta_N$, $\nu$) as obtained from 
scaling collapse analysis of the order parameters ($O_{N,B}$) 
for different $J_H$.
$L=24,32,40,48,64$ used here. 
Additional corroborating 
analysis with correlation ratios 
is shown in the Ref.~\cite{suppinfo}.
}
\begin{tabularx}{\textwidth}{*{10}{>{\centering\arraybackslash}X}}
\toprule
$J_H$ &  $\nu_{N}$ & $\nu_{V}$  & $\eta_{N}$ & $\eta_{V}$ & $g_{cN}$ & $g_{cV}$ & $\chi^2_N$ & $\chi^2_V$ \\
\toprule
0.0 & 0.53(3) & 0.63(1) & 0.44(5) & 0.49(2) & 0.168(1) & 0.167(1) & 1.08-1.68 & 1.69-2.46\\
0.01 & 0.45(2) & 0.54(3) & 0.23(3) & 0.42(4) & 0.174(1) & 0.171(1) & 1.19-1.63 & 1.38-1.73 \\
0.025 &  0.43(3) & 0.46(4) & 0.15(9) & 0.38(2)  & 0.182(1) & 0.180(1) & 0.75-1.46 & 0.8-1.4\\
0.04 & 0.40(2) & 0.43(5) & 0.13(7) & 0.30(8)  & 0.19(1) & 0.189(1) & 1.06-1.67 & 1.09-1.5\\
0.05 & 0.39(4) & 0.38(5) & 0.20(9) & 0.29(6)  & 0.196(1) & 0.195(1) & 0.87-1.31 & 0.87-1.96\\
0.07 & 0.38(2) & 0.39(3) & 0.10(4) & 0.10(4)  & 0.207(1) & 0.206(1) & 1.52-2.54 & 1.04-1.77\\
0.1 &  0.35(4) & 0.35(3) & -0.03(5) & -0.03(2)  & 0.224(1) & 0.224(1) & 1.24-3.28 & 0.99-1.97\\
0.15 &  0.33(2) & 0.33(1) & 0.00(8) & -0.12(8)  & 0.253(1) & 0.253(1) & 1.42-1.79 & 1.15-1.63\\
\toprule
\end{tabularx}
\label{dynamic}
\end{table*}

The first thing of note is that the anomalous exponents $\eta_N$ and
$\eta_V$ are markedly different as soon as $J_H \neq 0$.
For $J_H = 0$, we obtain $SU(3)$ exponents in overall agreement 
with earlier work~\cite{Lou_Sandvik_Kawashima_PRB_2009}
though our best estimate for the N\'eel correlation exponent $\nu_N$ 
is different.
The collapse quality when $\nu_N$ is set same as $\nu_V$ ($\sim 0.63$) is not
significantly worse. The equality $\nu_N = \nu_V$ is 
expected in the theory of DQC.
This marked difference of $\eta_N, \eta_V$
from $J_H=0$ suggests that $SU(3)$ criticality 
is not obtained when $J_H \neq 0$,
i.e. the $SU(2)$ perturbation changes the universality class. 
This is not entirely unexpected and can be taken as evidence for the 
doubled spin-$1/2$ DQC scenario~\cite{Wang_Kivelson_Lee_NatPhys_2015}.
However, there is a slow drift in the exponents as $J_H$ increases
which argues against a stable set of 
exponents~\cite{exponent_stability_footnote} 
as expected in the second scenario.
The drift is noticeable even within the accuracy levels
achieved by us. This level of accuracy
in the estimation of critical exponents is not unusual
in numerical studies of DQC. Similarly, the best estimates of
$\nu_N$ and $\nu_V$ for each value of $J_H$ do not match in all
cases, but again setting them equal does not lead to significant
loss of collapse quality. 
Nevertheless, we certainly see
that the anomalous exponents $\eta_N$, $\eta_V$ 
are not small. This is one of the 
expectations in the theory of DQC which is 
strikingly different than conventional
second-order critical points, and seen in 
previous $S=\frac{1}{2}$ studies.
Eventually for large enough $J_H=0.1$ and $0.15$, 
the anomalous exponents go negative indicating first-order behavior.
The correlation exponents $\nu_N$, $\nu_V \sim \frac{1}{2+1}$ for these
$J_H$ values as well.
This is to be expected when $J_H$ becomes large enough
since first-order
behaviour has been seen for $J_B=0$~\cite{Wildeboer_etal_prb2020}.

\begin{figure}[t]
    \centering
    \includegraphics[width=0.7\linewidth]{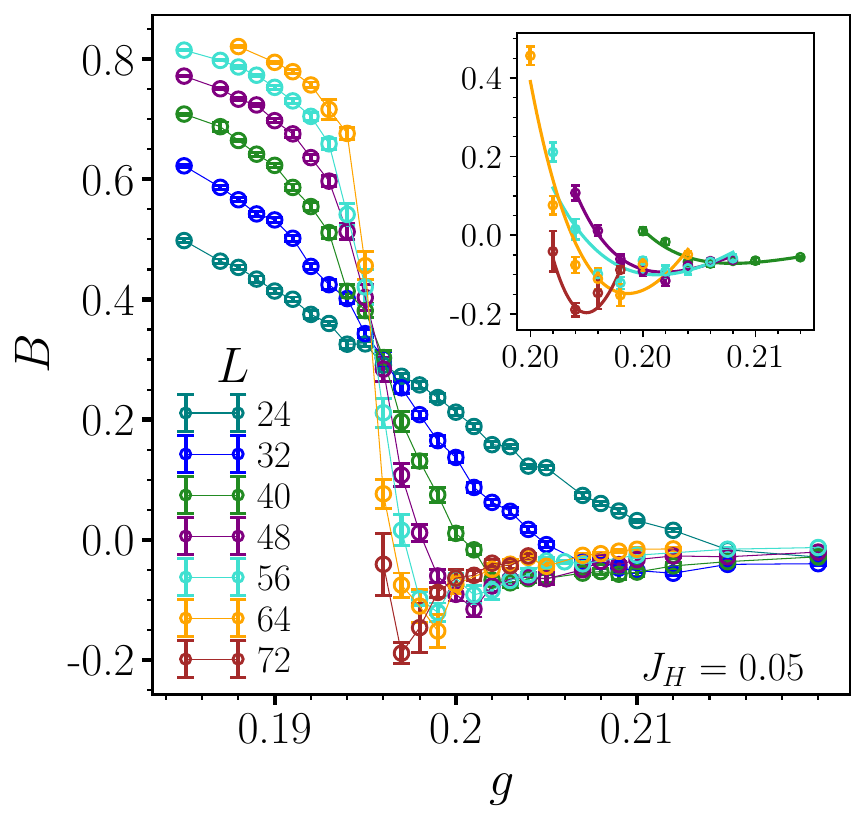}
    \caption{Binder ratio plotted versus fugacity $g=Q_{B}/J_{B}$ for $J_H=0.05$
    ($\beta=\frac{L}{4}$). Similar plots with
    negative dips are obtained for other $J_H$ values~\cite{suppinfo}. This
    calls into question the continuous nature
    of the transition. The inset 
    shows interpolations done in the vicinity of the Binder
    dip to extract an estimate of the dip
    magnitude for Fig.~\ref{fig:binderdips_vs_L}.
}
    \label{fig:binder_crossings}
\end{figure}

\begin{figure}
 \centering
 \includegraphics[width=0.9\linewidth]{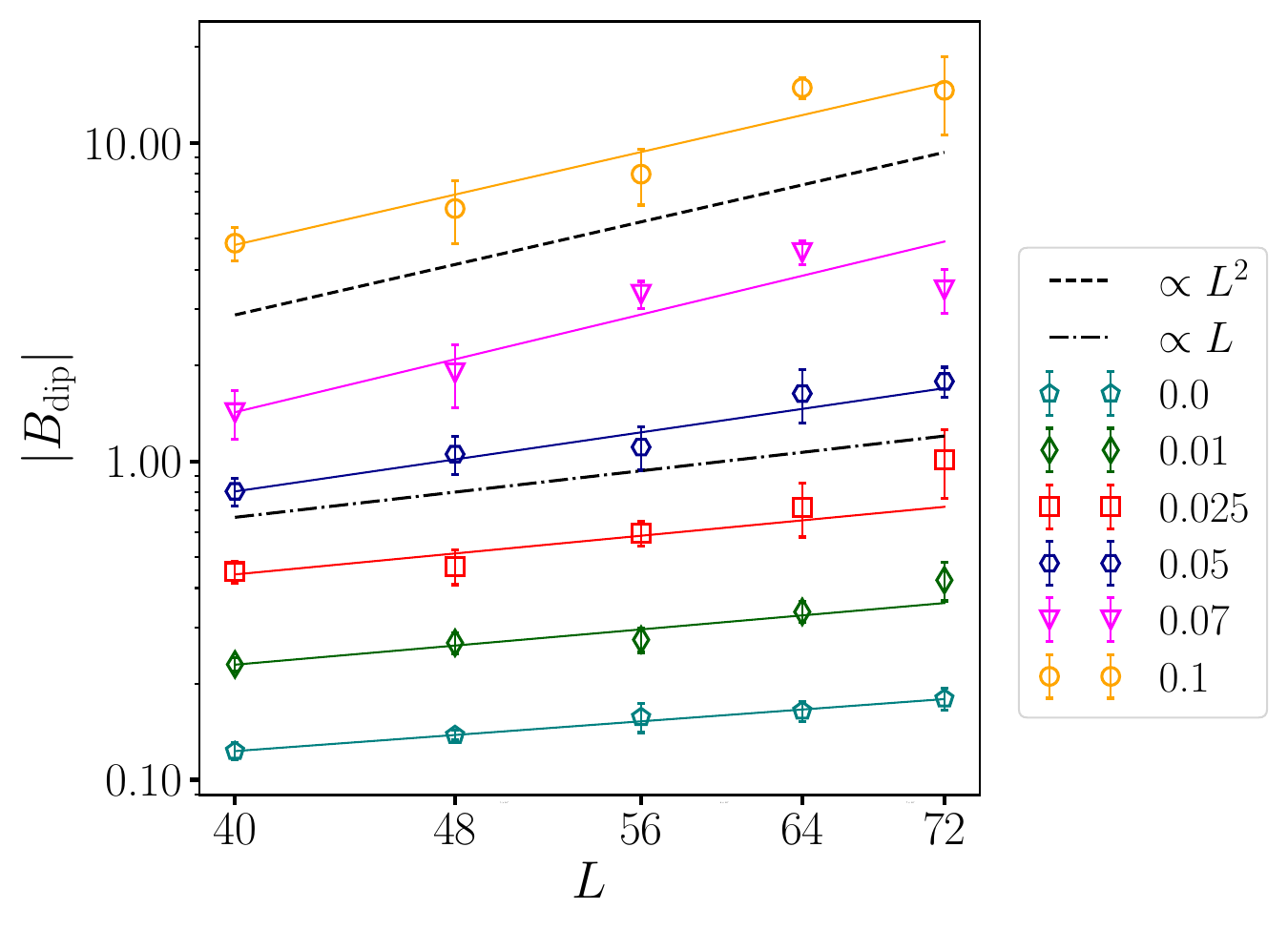}
 \caption{(Negative) Binder dip magnitude versus
 $L$ for different $J_H$ shown on log-log scale. 
It scales subextensively for small $J_H$.
We interpret this as 
(deconfined) pseudocriticality. 
The dashed lines are $L$ and
(extensive) $L^2$ power laws.
The different $J_H$ curves have been
scaled by different constant factors for clarity.
}
 \label{fig:binderdips_vs_L}
\end{figure}

From the preceding discussions, 
the N\'eel-cVBS transition in Eq.~\ref{eq:SU2Ham} appears continuous 
up to $J_H \sim 0.07$.
However, due to the observed drifts in the exponents, we further examine
the Binder ratios of the magnetization order parameter. 
We would expect them to
behave similar to the correlation ratios. 
Fig~\ref{fig:binder_crossings} shows this ratio, 
defined as 
$\frac{5}{2}(3-\frac{\langle m^4 \rangle}{\langle m^2 \rangle^2})$, 
with a clearly visible crossing at the transition point. 
Equally noteworthy is the clear dip below zero near the transition.
This is seen for all values of $J_H$~\cite{suppinfo}. 
A characteristic of first order transitions is that this dip 
grows extensively with system size~\cite{Vollmayr_etal_1993}. 
In Ref~\cite{Kaul_prb2011}, where the $J_H=0$ case was
studied, it had been noted that this dip grows 
sub-extensively with system size and interpreted as evidence for a 
continuous transition. We similarly find 
sub-extensively growing dips for small $J_H$
as shown in Fig~\ref{fig:binderdips_vs_L}. 
As $J_H$ grows 
larger, it eventually grows as $L^2$ as expected for a first order 
transition~\cite{Sen_Sandvik_PRB2010}
concomitantly with $\nu_N$, $\nu_V \sim \frac{1}{2+1}$.

Given the drifting exponents 
seen earlier that argues against a bonafide DQC scenario~\cite{Demidio_footnote},
we rather interpret the sub-extensive growing Binder dips 
along with the $U(1)$-symmetric VBS histograms
as evidence for deconfined pseudocriticality.
In other words, the scenario of a doubled $S=\frac{1}{2}$ DQC 
provides a framework to understand the above set of numerical results,
but the deconfinement gets curtailed at much larger length scales (dependent
on $J_H$)
than the lattice scale leading to the observed pseudocritical behavior. 
In terms of QMC simulations, we 
imagine it as pseudo-DQC ensembles occurring in both the 
split $S=\frac{1}{2}$ Hilbert spaces in (QMC) space-time which then gets 
``inherited" by the $S=1$ system under projection back to the
fully symmetric subspace.
This also implies a revision of the earlier interpretation of DQC 
in $SU(3)$ $JQ$ model~\cite{Kaul_prb2011}.

We 
situate our interpretation of deconfined pseudocriticality 
in our $S=1$ microscopic model 
in the light of recent developments that have thrown open 
the issue of the second-order nature of 
DQCs~\cite{demidio_etal_arxiv2021,Zhao_etal_PRL2022,yuan_etal_arxiv2023,Song_etal_Arxiv2023}. 
These arose in the context of earlier works 
regarding emergent symmetry expectations at these
transitions when interpreted in a $SO(5)$ framework with a combined
order parameter built out of the N\'eel and VBS order parameters
~\cite{Tanaka_Hu_prl2005,Senthil_Fisher_prb2006}.
This would imply an enhanced emergent symmetry between the two symmetry unrelated
order parameters. This basic expectation has been numerically studied in
various cases~\cite{nahum_etal_prl2015,nahum_etal_prx2015,Fuzzy_Sphere_SO5_arxiv}, in 
certain unconventional transitions~\cite{Zhao_natphys2019} 
including  in one dimension~\cite{patil_katz_sandvik_prb2018,Xi_2022} 
and classical dimer models~\cite{sreejith_prl2019}. System size 
restrictions however can make interpretation of emergent symmetry tricky. 
From the theory side, conformal
bootstrap results~\cite{Nakayama_Ohtsuki_PRL_2016}
pointed out strong constraints on the scaling dimensions
that apparently rule out emergent symmetry in $2+1d$ DQC.
The notion of DQC as pseudocriticality was thus conjectured~\cite{Ma_Wang_PRB_2020,Nahum_PRB_2020} 
in order to reconcile with the conformal bootstrap result.
The idea is that the renormalization group (RG) 
fixed point for DQC does not reside
in the (physical) $2+1d$, but slightly below it 
(see Fig.~1 of Ref.~\cite{Ma_Wang_PRB_2020}) such
that pseudocritically slow flows obtain near the critical point
$g_c$. 
For more discussion on these issues, see this recent
review~\cite{Senthil_review_2023}.
Such slow RG flows may account for 
the good scaling collapse seen in numerical data for 
accessible system sizes \emph{along} with the observed
drifts in the critical exponents.
Recent quantum entanglement based 
results~\cite{Zhao_etal_PRL2022,yuan_etal_arxiv2023,Song_etal_Arxiv2023} 
have taken forward a similar line of argument for $SU(N)$ models
in general,
however another work~\cite{Demidio2024entanglement} 
provides a counterargument.

It is noteworthy that even before the emergent symmetry point
of view gained currency, numerical studies of DQC had seen
anomalous scaling corrections whose origin was unclear. 
Ref.~\onlinecite{Bartosch_PRB_2013} gave an explanation based on scaling corrections
inherent in the effective $U(1)$ gauge theory of the 
deconfined spinons.
Another explanation based on two different 
length scales
associated with spinon correlations and VBS domain wall sizes
diverging with different exponents
has also been proposed~\cite{Shao_Guo_Sandvik_Science2016}.
In the context of our interpretation, pseudocriticality 
indicates at the confinement of the putative spinons
of the $U(1)$ gauge theory framework for all values of the tuning
parameter as mentioned previously. 
Near the transition, the confinement length scale 
must become very large
($L \gtrsim O(100)$ given the high quality of scaling collapses
seen) compared to the lattice scale~\cite{Sandvik_youtube_2022} but
remain finite due to
pseudocritical nature of the RG flows.
The present consensus seems to be veering towards 
deconfined pseudocriticality
based on recent $S=\frac{1}{2}$ results, and our work
provides a microscopic $S=1$ model for this
scenario. 
This opens a question in the context of $SU(N)$ DQC. 
On one hand, theoretical expectations based on the suppression
of gauge fluctuations with increasing $N$ make the case for
bonafide DQC. On the other hand, no negative Binder dips
has been numerically seen in $SU(2)$ $JQ$ models for 
largest sizes studied.
This makes the pattern of Binder dip growth with respect to
$N$ mysteriously non-monotonic~\cite{fig1_Kaulprb2011_footnote}.
It will be an useful theoretical advance and 
strong evidence for the pseudocriticality scenario if the 
sub-extensive growth of negative Binder ratio dips can be 
linked to the pseudocritcal RG
flows of Refs.~\cite{Nahum_PRB_2020,Ma_Wang_PRB_2020}.

\prlsec{Acknowledgements}
We acknowledge useful discussions with Fabien Alet, Subhro Bhattacharjee,
Kedar Damle, Prashant Kumar and Adam Nahum.
VVi was supported by the institute post-doctoral
fellowship program at IIT Bombay,
and in part by the International Centre for Theoretical Sciences (ICTS) 
during the program - 8th Indian Statistical Physics 
Community Meeting (code: ICTS/ISPCM2023/02).
SP acknowledges funding support from SERB-DST, India 
via Grants No. SRG/2019/001419 and MTR/2022/000386.
Partial support by Grant No. CRG/2021/003024 is also
acknowledged.
ND was initially supported by National Postdoctoral Fellowship of 
SERB, DST, Govt. of India (PDF/2020/001658) at the department of 
Theoretical Physics, TIFR and presently by the TIFR postdoctoral 
fellowship.
The numerical results were obtained using
the computational facilities of the Department of Physics,
Indian Institute of Technology (IIT) Bombay.

\bibliographystyle{apsrev}
\bibliography{Seq1_refs}

\begin{thebibliography}{92}
\bibitem[{cor()}]{corr_ratio_footnote}
\bibinfo{note}{These are calculated by taking the ratio of the Fourier
  transform of the correlation functions ($\tilde{C}(k)$) at the ordering
  momentum, $k_{\text{o}}$, and a momentum closest to it (in some chosen
  direction) at the given system size, $k_{\text{o}'}$, $R \equiv
  1-\frac{\tilde{C}(k_{\text{o}'})}{\tilde{C}(k_{\text{o}})}$. It can be
  verified that $R \rightarrow 1$ and 0 as $L\rightarrow \infty$ in the ordered
  and disordered phases respectively.}
\bibitem[{cvb()}]{cvbs_footnote}
\bibinfo{note}{Note we use the biquadratic exchange term to construct the cVBS
  order parameter instead of the more frequently used Heisenberg exchange term.
  Both constructions can probe cVBS order; our choice has better statistics in
  terms of QMC estimation for the parameter regime we are interested in, i.e.
  the vicinity of $J_H=0$.}
\end{thebibliography}


\end{document}


\title{Supplementary information for ``Deconfined pseudocriticality in a 
model spin-1 quantum antiferromagnet"}

\author{Vikas Vijigiri}
\altaffiliation[vikas.v@iitb.ac.in]{}\affiliation{Department of Physics, Indian Institute of Technology Bombay, Powai, Mumbai, 400076, India}
\author{Nisheeta Desai}
\affiliation{Department of Theoretical Physics, Tata Institute of Fundamental Research,  Mumbai, 400005, India}
\author{Sumiran Pujari}
\altaffiliation[sumiran@phy.iitb.ac.in]{}\affiliation{Department of Physics, Indian Institute of Technology Bombay, Powai, Mumbai, 400076, India}

\date{\today}

\begin{abstract}
Here we present additional data and plots to  
document the data sets collected during this work
for the various Heisenberg exchange coupling ($J_H$) values 
studied in the main text.
\end{abstract}

\maketitle
\tableofcontents

\section{Model and Observables}
\label{sec:model}
We recapitulate again the various definitions
related to the model and observables introduced in the main text 
for convenience.
The $S=1$ Hamiltonian (Eq.~1 and Eq.~2 of the main text) is given by,
\begin{eqnarray}
    H &=& J_H\sum_{\langle ij \rangle} 
    \{ \mathbf{S}_i\cdot \mathbf{S}_j - \mathbb{I} \}
    - 
    \frac{J_B}{3} \sum_{\langle ij \rangle} 
    \{(\mathbf{S}_i\cdot \mathbf{S}_j)^2  - \mathbb{I} \}
    \nonumber \\
    & & - \; \frac{Q_B}{9} \sum_{\langle ijkl \rangle} 
    \left[ \{(\mathbf{S}_i\cdot \mathbf{S}_j)^2 - \mathbb{I}\}
    \{(\mathbf{S}_k\cdot \mathbf{S}_l)^2 - \mathbb{I}\}
    +
    \{(\mathbf{S}_i\cdot \mathbf{S}_l)^2 - \mathbb{I}\}
    \{(\mathbf{S}_j\cdot \mathbf{S}_k)^2 - \mathbb{I}\}
    \right]
\end{eqnarray}
where $J_H, J_B$ are the coefficients of the Heisenberg and Biquadratic 
exchange terms, and $Q_B$ is the coefficient of the designer $Q$-term composed
of biquadratic exchanges respectively. 

  The following structure factors  
  \begin{eqnarray}
  \mathcal{A}(\mathbf{q}) &=&\frac{1}{N^2} \sum\limits_{\mathbf{r},\mathbf{r}^\prime} e^{i(\mathbf{r}-\mathbf{r}^\prime)\cdot \mathbf{q}}\langle S^z_{\mathbf{r}} S^z_{\mathbf{r}^\prime} \rangle \\   
  \mathcal{B}(\mathbf{q}) &=& \frac{1}{N^2} \sum\limits_{\mathbf{r},\mathbf{r}^\prime} e^{i(\mathbf{r}-\mathbf{r}^\prime)\cdot \mathbf{q}}
  \langle \left( \mathbf{S}_{\mathbf{r}}\cdot \mathbf{S}_{\mathbf{r}^\prime} -\mathbb{I} \right) \rangle  \\  
  \mathcal{C}^{\mu}(\mathbf{q}) &=& \frac{1}{N^2} \sum\limits_{\mathbf{r},\mathbf{r}^\prime} e^{i(\mathbf{r}-\mathbf{r}^\prime)\cdot \mathbf{q}}\langle 
  (\mathbf{S}_{\mathbf{r}}\cdot \mathbf{S}_{\mathbf{r}+\mathbf{e}_\mu}-\mathbb{I}) 
  \times
 (\mathbf{S}_{\mathbf{r}^\prime} \cdot \mathbf{S}_{\mathbf{r}^\prime+\mathbf{e}_\mu}-\mathbb{I}) \rangle  \\
  \mathcal{D}^\mu(\mathbf{q}) &=& \frac{1}{9N^2} \sum\limits_{\mathbf{r},\mathbf{r}^\prime} e^{i(\mathbf{r}-\mathbf{r}^\prime)\cdot \mathbf{q}}\langle ((\mathbf{S}_{\mathbf{r}}\cdot \mathbf{S}_{\mathbf{r}+\mathbf{e}_\mu})^2-\mathbb{I}) \times
  ((\mathbf{S}_{\mathbf{r}^\prime}\cdot \mathbf{S}_{\mathbf{r}^\prime+\mathbf{e}_\mu})^2-\mathbb{I})) \rangle 
  \end{eqnarray}
  made out of real-space correlators can serve as 
  order parameters for N\'eel and VBS ordering.
As a side remark, $\mathcal{B}(\mathbf{q})$ can be measured during loop update
in SSE. However, we do not need to do this as $\mathcal{A}(\mathbf{q})$ 
which also detects N\'eel ordering 
can measured in  a much simpler way. Furthermore, in presence of 
$SU(2)$ symmetry they are linearly related to each other.
Therefore, we focused on $\mathcal{A}(\mathbf{q})$ to probe N\'eel order
at the antiferromagnetic ordering wavevector on the square lattice
($(\pi,\pi)$ when lattice constants are set to unity)
as mentioned in the main text as well.
The corresponding N\'eel correlation ratio $R_N$ is defined to be
\begin{eqnarray}
R_N \equiv \frac{R_x + R_y}{2}
\end{eqnarray}
where,
\begin{eqnarray}
R_x & \equiv& 1 - \frac{ \mathcal{A}(\pi + \frac{2\pi}{L},\pi)}
{\mathcal{A}(\pi,\pi)}  \\
R_y &\equiv& 1 - \frac{\mathcal{A}(\pi, \pi + \frac{2\pi}{L})}
{\mathcal{A}(\pi,\pi)}
\end{eqnarray}

Similarly, both $\mathcal{C}(\mathbf{q})$ and $\mathcal{D}(\mathbf{q})$
made out of bond-bond correlators of Heisenberg and biquadratic couplings
can be used to probe VBS order at its ordering wavevector on the square lattice
($(\pi,0)$ and $(0,\pi)$ when lattice constants are set to unity).
Either one would thus suffice for the study of N\'eel-VBS transitions.
The Heisenberg bond energy based VBS correlation ratio $R_{V'}$ is 
defined to be
\begin{eqnarray}
R_{V'} \equiv \frac{R_{V'_x} + R_{V'_y}}{2}
\end{eqnarray}
where, 
\begin{eqnarray}
R_{V'_x} & \equiv & 1 - \frac{\mathcal{C}(\pi + \frac{2\pi}{L},0)}
{\mathcal{C}(\pi,0)} \\
R_{V'_y} & \equiv & 1 - \frac{\mathcal{C}(0, \pi + \frac{2\pi}{L})}
{\mathcal{C}(0,\pi)}
\end{eqnarray}
We have rather focused on the biquadratic bond energy based VBS 
observables in the paper due to better statistics while estimating
it during QMC compared to the Heisenberg bond energy based observables
as discussed also in a footnote in the main text.
The corresponding VBS correlation ratio is thus defined to
\begin{eqnarray}
R_V \equiv \frac{R_{V_x} + R_{V_y}}{2}
\end{eqnarray}
where, 
\begin{eqnarray}
R_{V_x} &=& 1 - \frac{\mathcal{D}(\pi + \frac{2\pi}{L},0)}{D(\pi,0)} \\
R_{V_y} &=& 1 - \frac{\mathcal{D}(0, \pi + \frac{2\pi}{L})}{\mathcal{D}(0,\pi)}
\end{eqnarray}
Finally, from the above we can also see the correspondence between the notation
of the main text and that used here for the order parameters as
\begin{eqnarray}
 O_N & = & \mathcal{A}(\pi,\pi)   \\
 O_V & = & \mathcal{D}^x(\pi,0) + \mathcal{D}^y(0,\pi)
\end{eqnarray}

\clearpage
\pagebreak[4]

\section{Data sets}
\label{sec:datasets}

\subsection{Convergence with inverse temperature $\beta$}
\label{subsec:beta_convergence}

\begin{figure*}[h!]
    \centering
\includegraphics[width=0.9\linewidth]{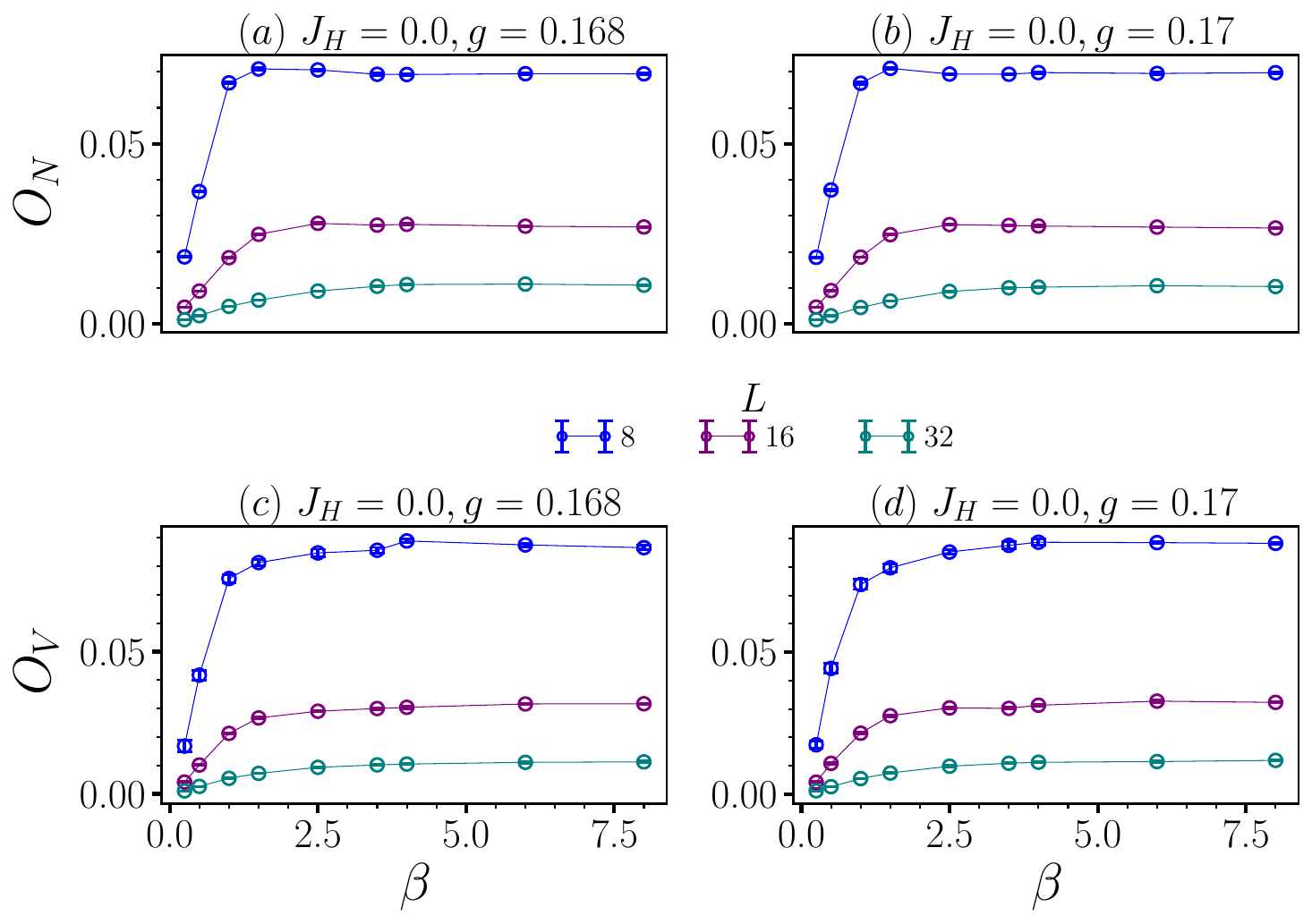}   
    \caption{N\'eel and cVBS order parameters ($O_{N}$
    and $O_V$) versus the inverse temperature $\beta=\frac{1}{T}$.  
    $\frac{J_H}{J_B}=0.0$,  $g \equiv \frac{Q_B}{J_B} \sim 0.17$.
    We recall that $J_B$ is set to 1 everywhere.
    Similar behaviour is present everywhere in the parameter regimes
    explored as illustrated through the next two figures. 
    We see that $\beta = \frac{L}{4}$ is more than sufficient
    for making conclusions about ground state properties.}
    \label{order_param}
\end{figure*}

\begin{figure*}[h!]
    \centering
\includegraphics[width=0.9\linewidth]{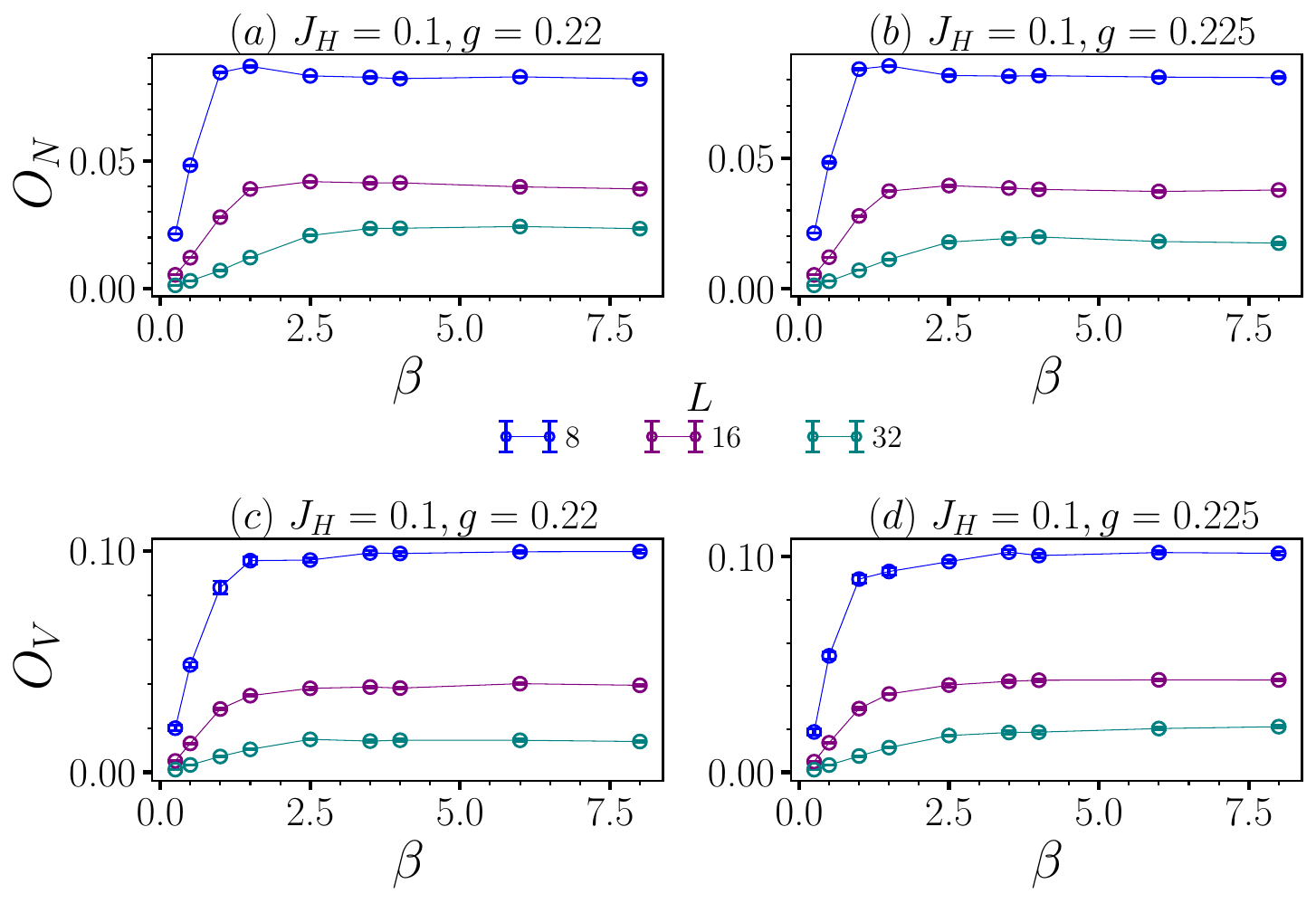}      
    \caption{N\'eel and cVBS order parameters ($O_{N}$
    and $O_V$) versus the inverse temperature $\beta=\frac{1}{T}$.  
    $\frac{J_H}{J_B}=0.1$,  $g \equiv \frac{Q_B}{J_B} \sim 0.22$.
    }
    \label{order_param}
\end{figure*}

\begin{figure*}[h!]
    \centering
\includegraphics[width=0.45\linewidth]{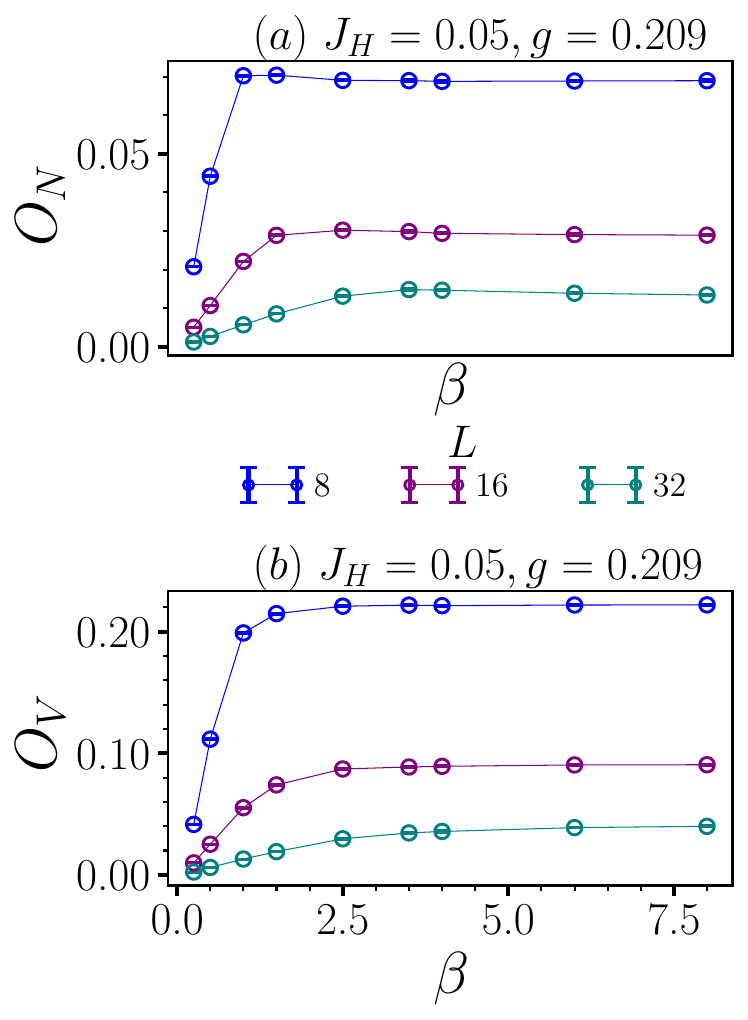}   
    \caption{N\'eel and cVBS order parameters ($O_{N}$
    and $O_V$) versus the inverse temperature $\beta=\frac{1}{T}$.  
    $\frac{J_H}{J_B}=0.0$,  $g \equiv \frac{Q_B}{J_B} \sim 0.21$.
    }
    \label{order_param}
\end{figure*}

\clearpage
\pagebreak[4]

\subsection{Correlation Ratios}
\label{subsec:corr_ratio}

\begin{figure*}[h!]
    \centering
    \includegraphics[width=16cm,height=17.5cm]{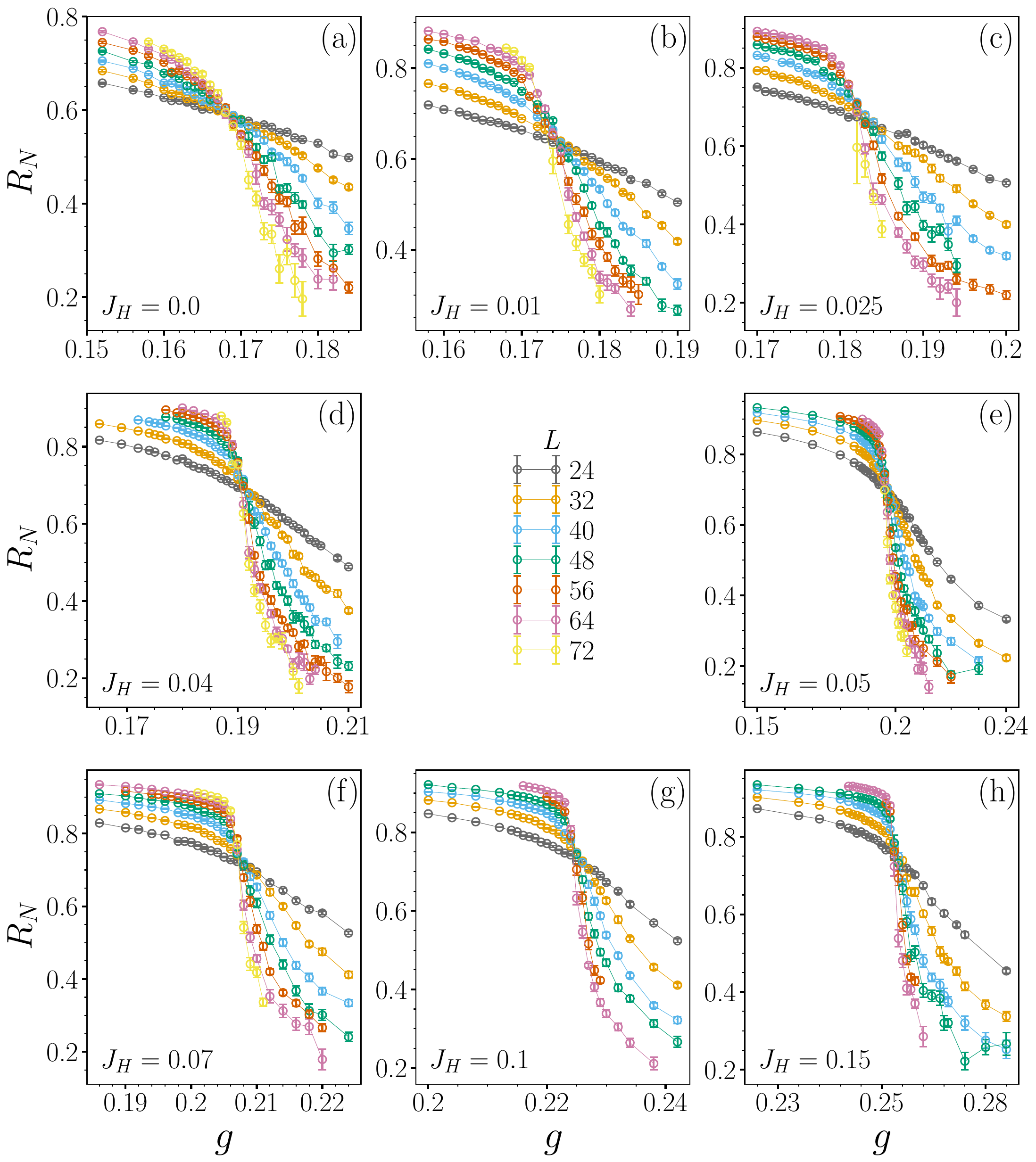}    
    \caption{\label{corr_ratio}
    This figure expands on Fig.~1a of the main text to document all the $R_N$ 
    correlation ratio data sets collected.
    $\beta=\frac{L}{4}$ throughout.
    }
\end{figure*}
\begin{figure*}[h!]
    \centering
    \includegraphics[width=16cm,height=17.5cm]{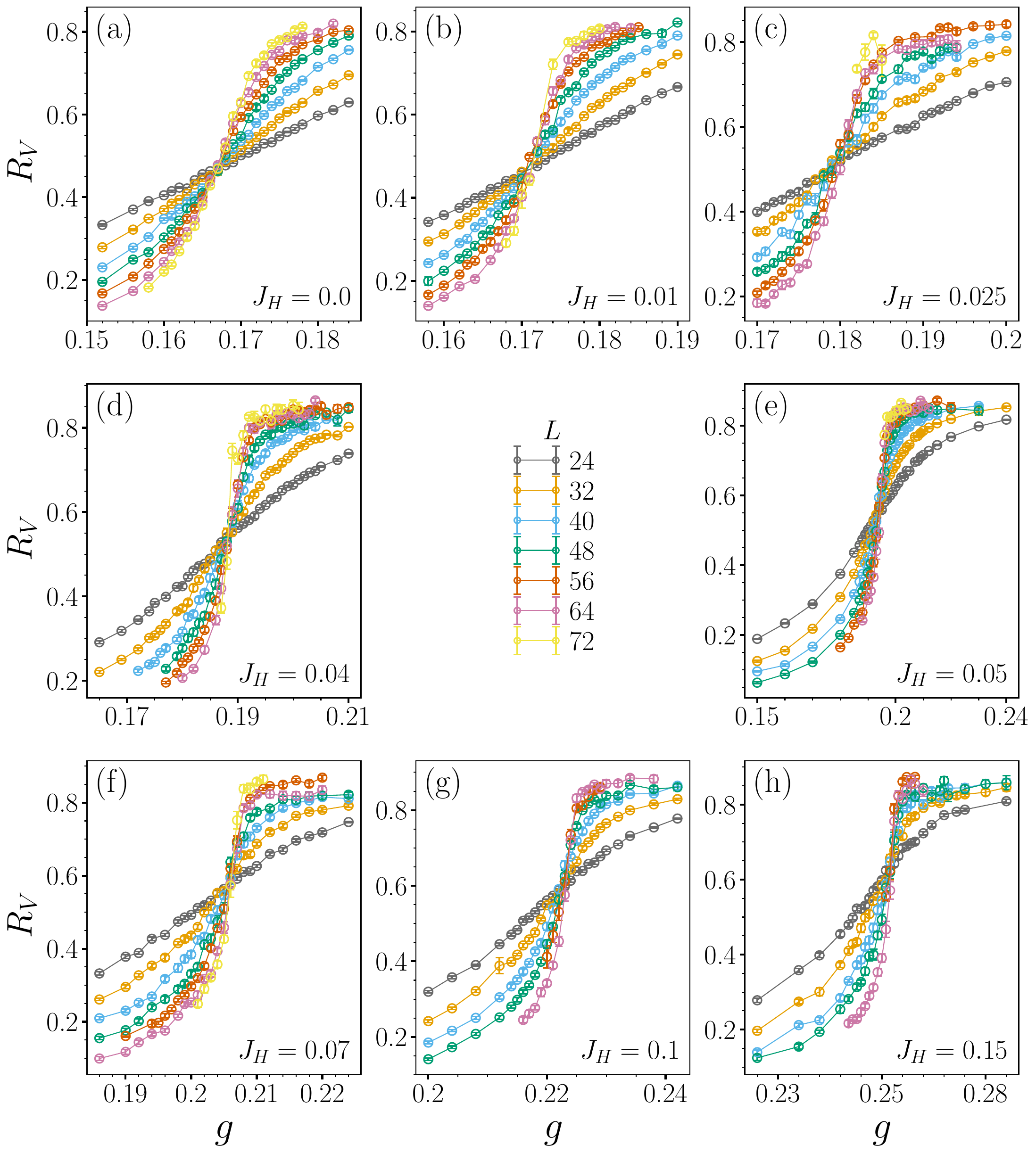}    
    \caption{ \label{corr_ratio}
   This figure expands on Fig.~1b of the main text to document all the $R_V$ 
   correlation ratio data sets collected.
    $\beta=\frac{L}{4}$ throughout.
    }
\end{figure*}

\clearpage
\pagebreak[4]
\subsection{Staggered magnetization histograms}
\begin{figure*}[h!]
    \centering
    \includegraphics[width=0.825\linewidth]{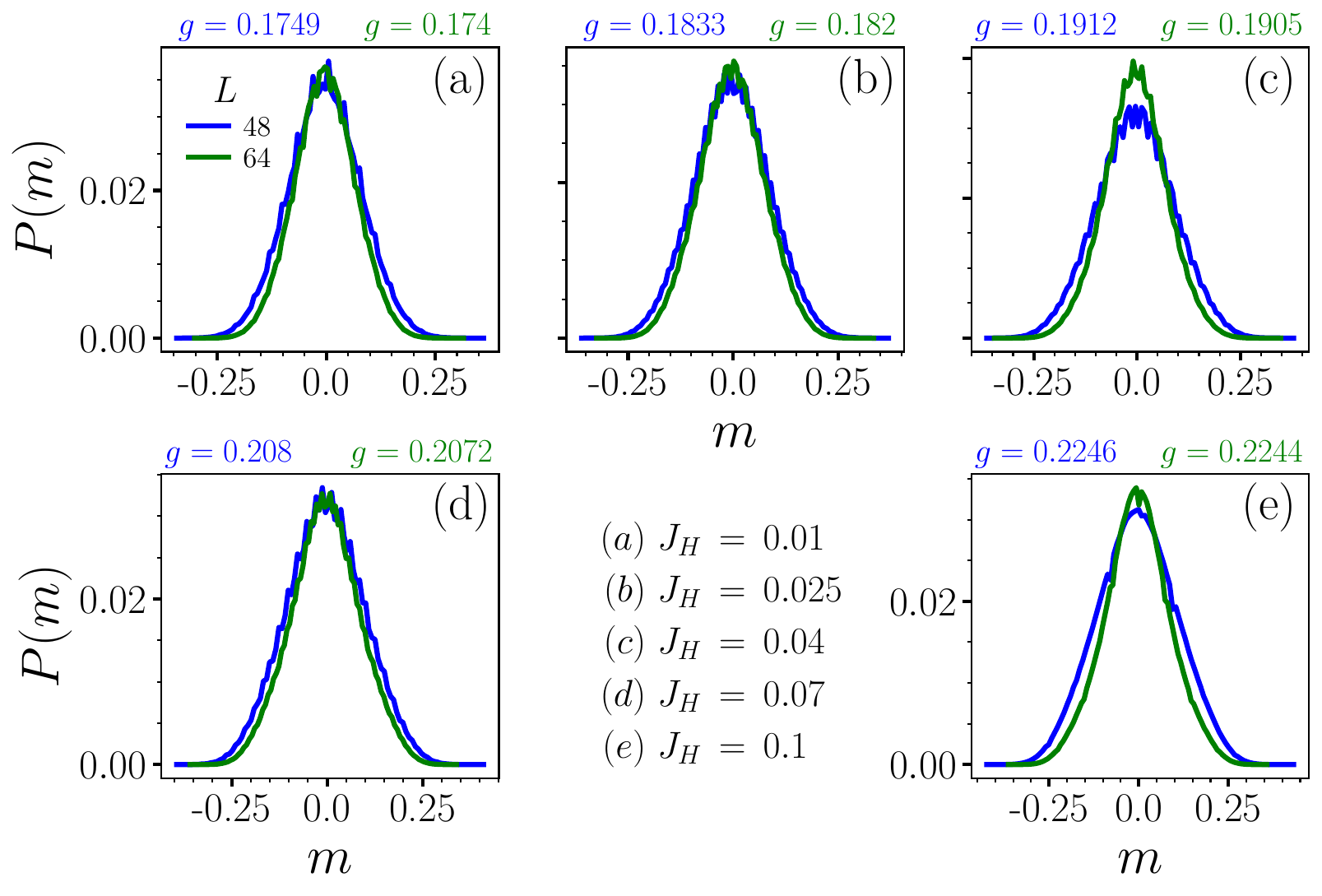}  
\includegraphics[width=0.825\linewidth]{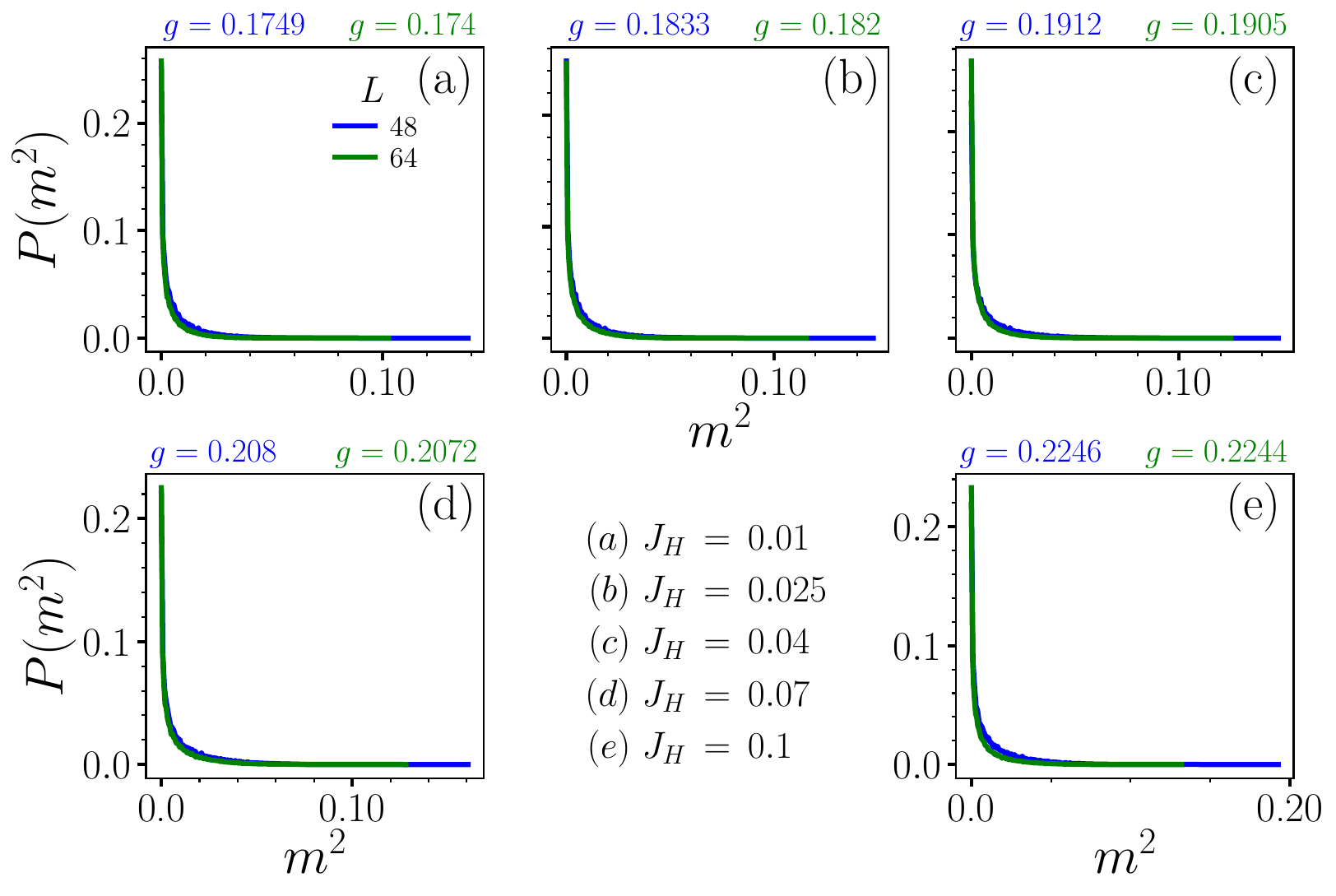}    
    \caption{\label{binder_plot}
    This figure expands on Fig.~2a of the main text to document the various
    staggered magnetization histogram data collected.
     $\beta=\frac{L}{4}$ throughout. The $g$ values above correspond
     to $\sim g_c(L)$.
    }
\end{figure*}

\clearpage
\pagebreak[4]
\subsection{cVBS order parameter histograms}
\begin{figure*}[h!]
    \centering
\includegraphics[width=0.825\linewidth]  {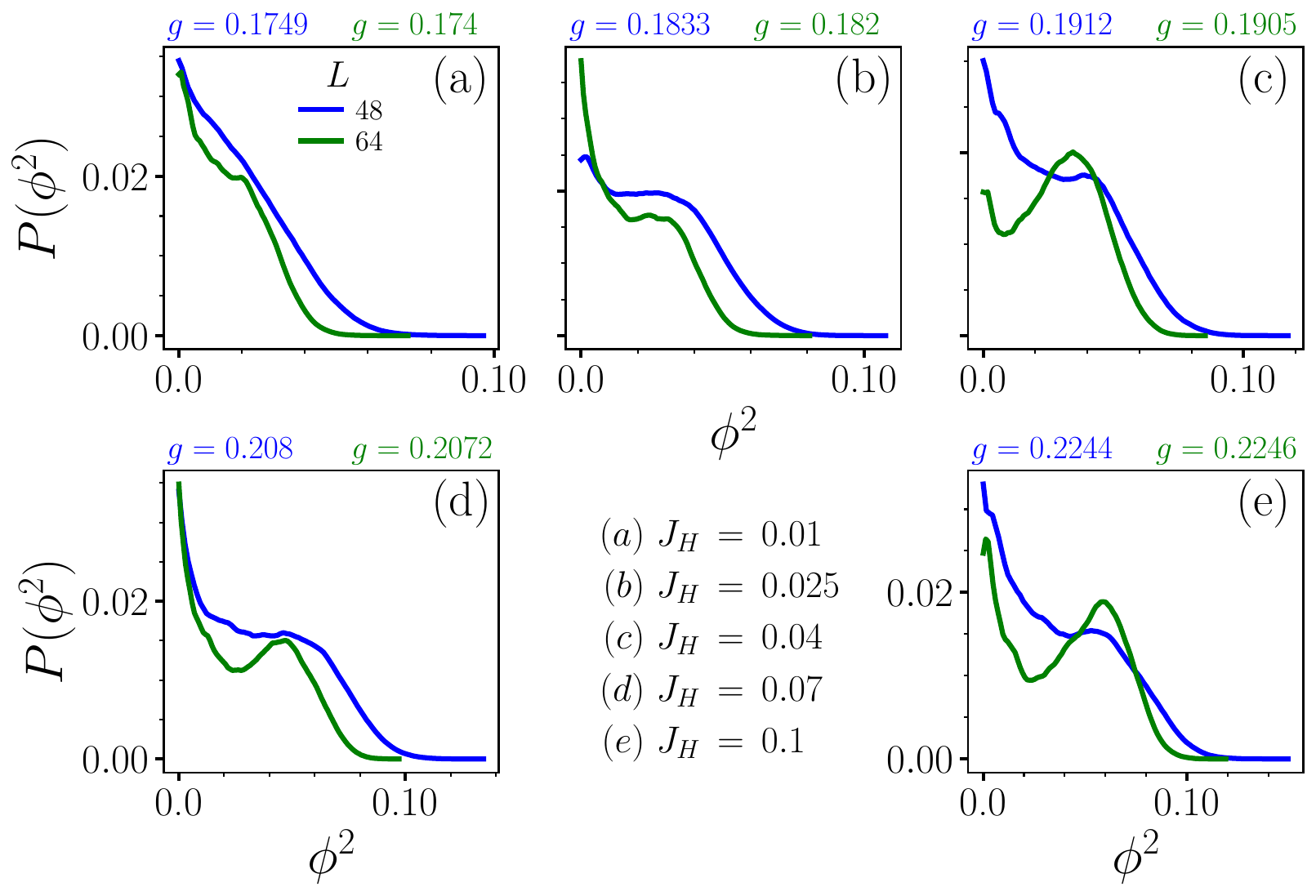}
\includegraphics[width=0.825\linewidth]{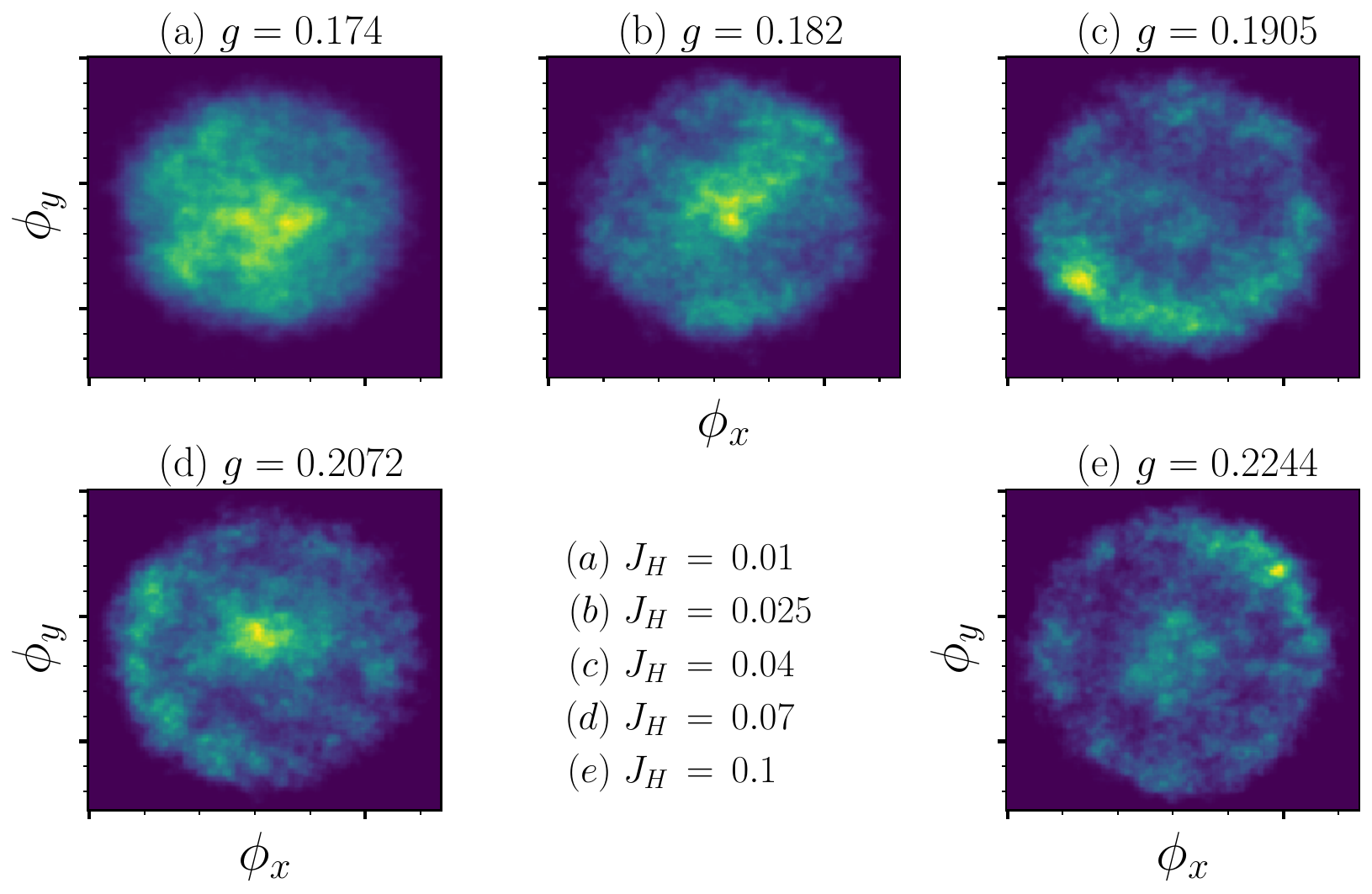} 
    \caption{\label{binder_plot}
    This figure expands on Fig.~2b of the main text to document the various
    cVBS histogram data collected.
     $\beta=\frac{L}{4}$ throughout. $L=64$ for cVBS $(\phi_x,\phi_y)$ heat maps.
     The $g$ values above correspond to $\sim g_c(L)$. 
    }
\end{figure*}

\clearpage
\pagebreak[4]
\begin{figure*}[h!]
    \centering
\includegraphics[width=0.825\linewidth]{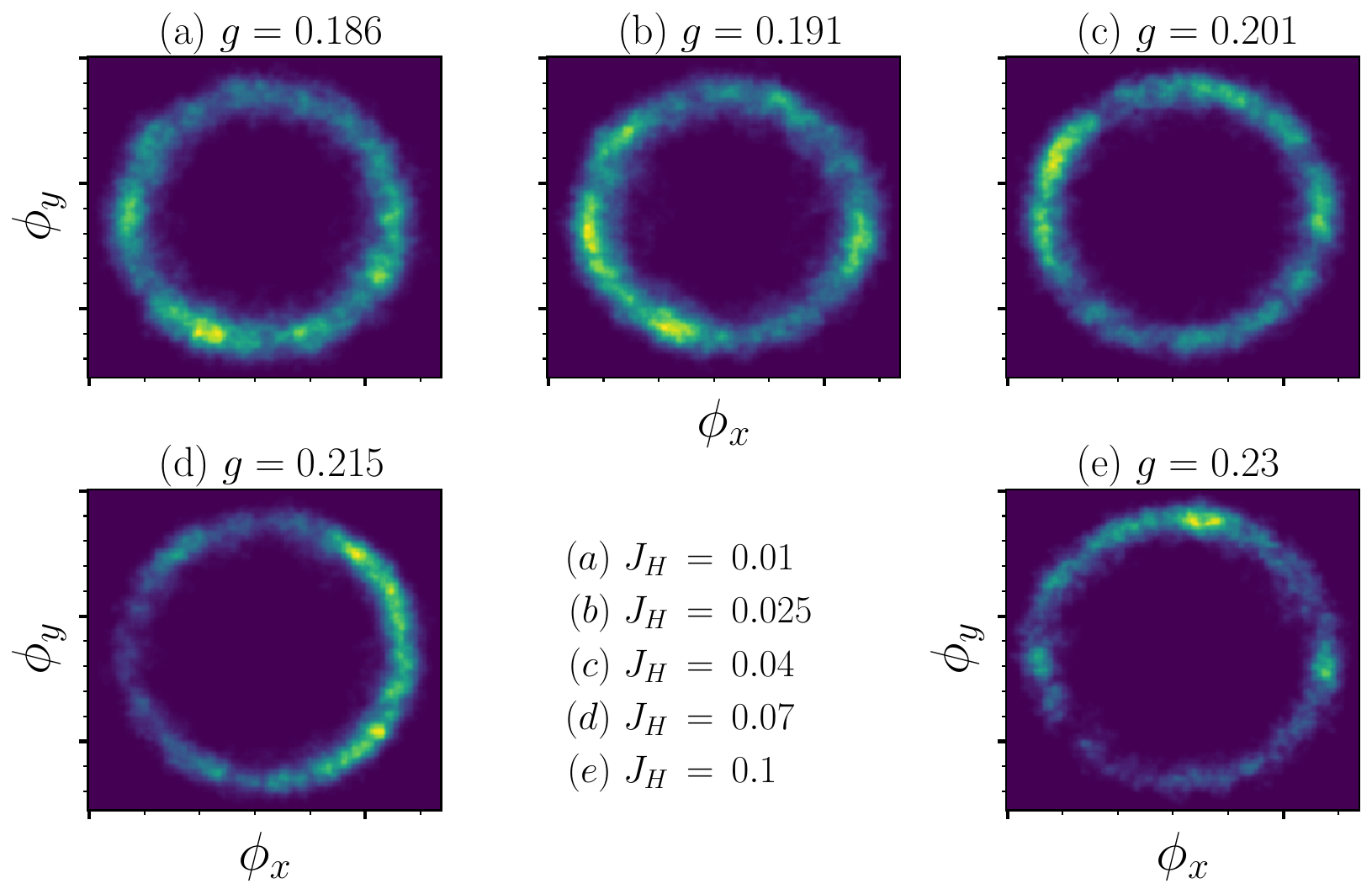} 
    \caption{\label{binder_plot}
     cVBS $(\phi_x,\phi_y)$ heat maps
     for $g$ slightly to the right of $g_C(L)$ on
     the VBS side to better highlight
     the ``$U(1)$-symmetric" nature of these histograms near
     the phase transition.
     $\beta=\frac{L}{4}$ and $L=64$ throughout. 
    }
\end{figure*}

\clearpage
\pagebreak[4]

\subsection{Staggered magnetization time series}
\begin{figure*}[h!]
    \centering
     \includegraphics[width=0.825\linewidth]{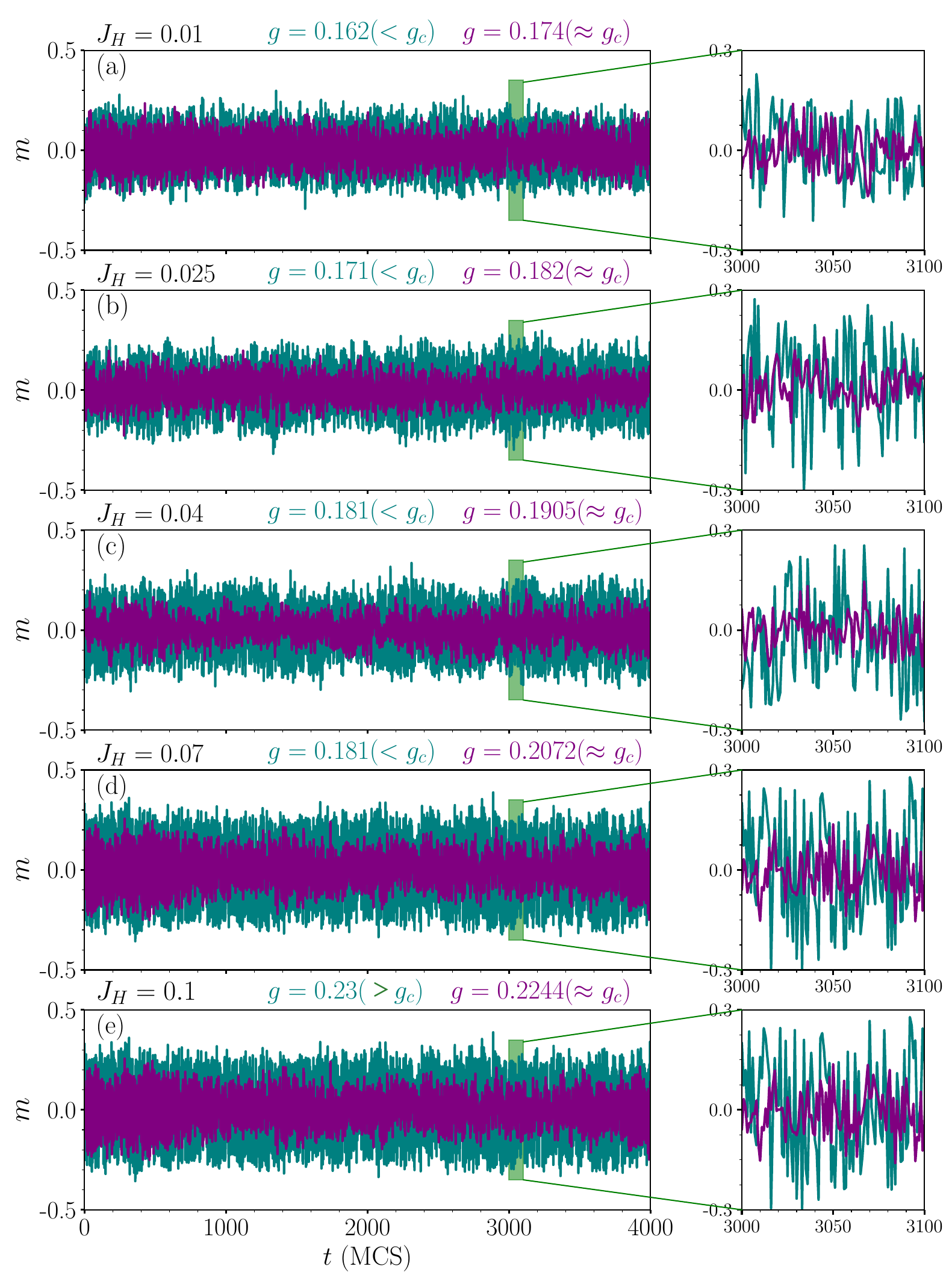}   
   
    \caption{\label{m_timeseries}
    This figure expands on Fig.~2c to document the various time series 
    data collected on staggered magnetization.
     $\beta=\frac{L}{4}$ and $L=64$ throughout. 
     The time series data
     are in register with those in Fig.~\ref{m2_timeseries}.
    }
\end{figure*}

\clearpage
\pagebreak[4]

\begin{figure*}[h!]
    \centering
     \includegraphics[width=0.825\linewidth]{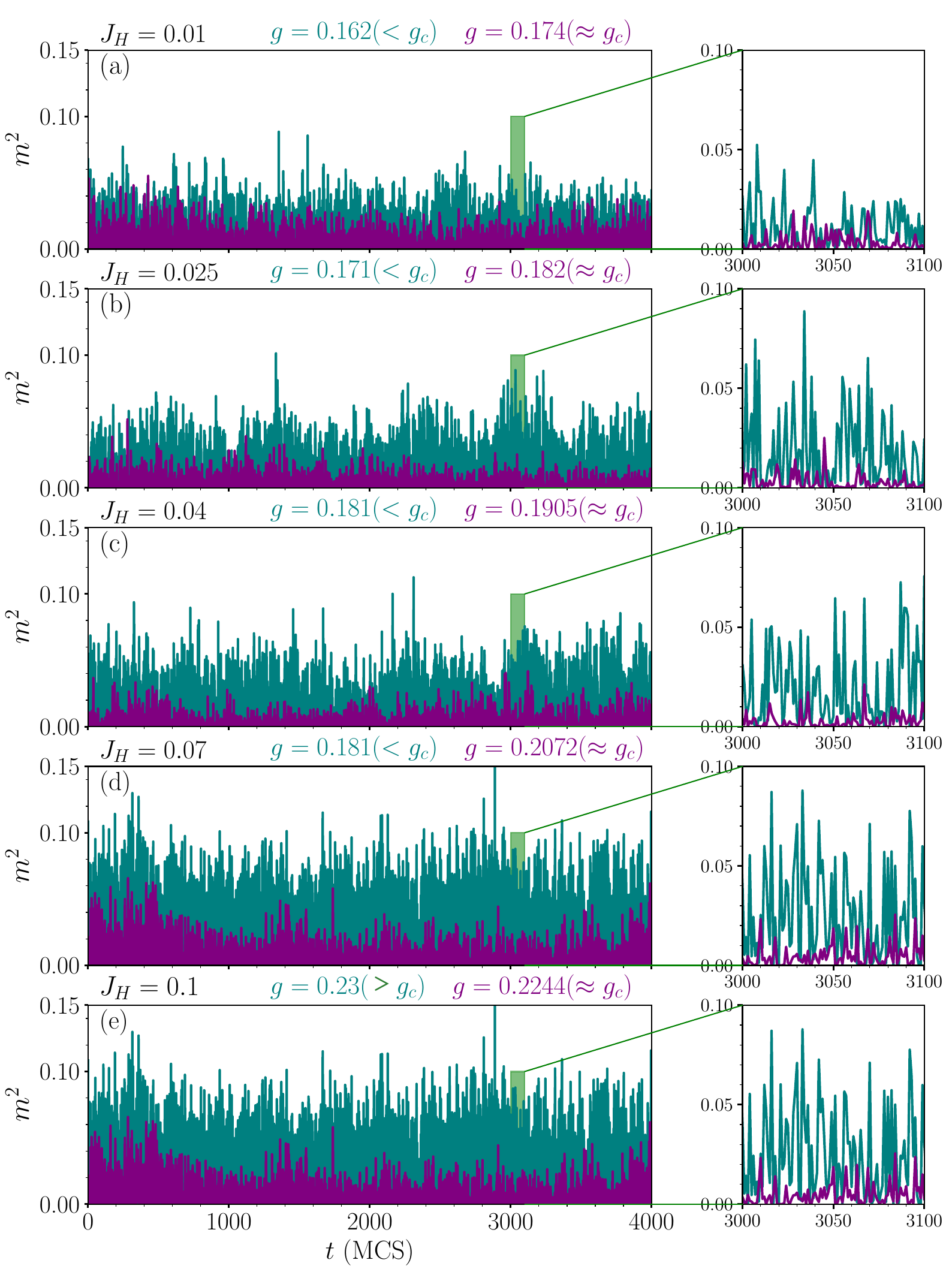}   
   
    \caption{\label{m2_timeseries}
    This figure expands on Fig.~2c to document the various time series 
    data collected on staggered magnetization.
     $\beta=\frac{L}{4}$ and $L=64$ throughout. 
     The time series data
     are in register with those in Fig.~\ref{m_timeseries}.
    }
\end{figure*}

\clearpage
\pagebreak[4]

\subsection{cVBS order parameter time series}
\begin{figure*}[h!]
    \centering
    \includegraphics[width=0.825\linewidth]{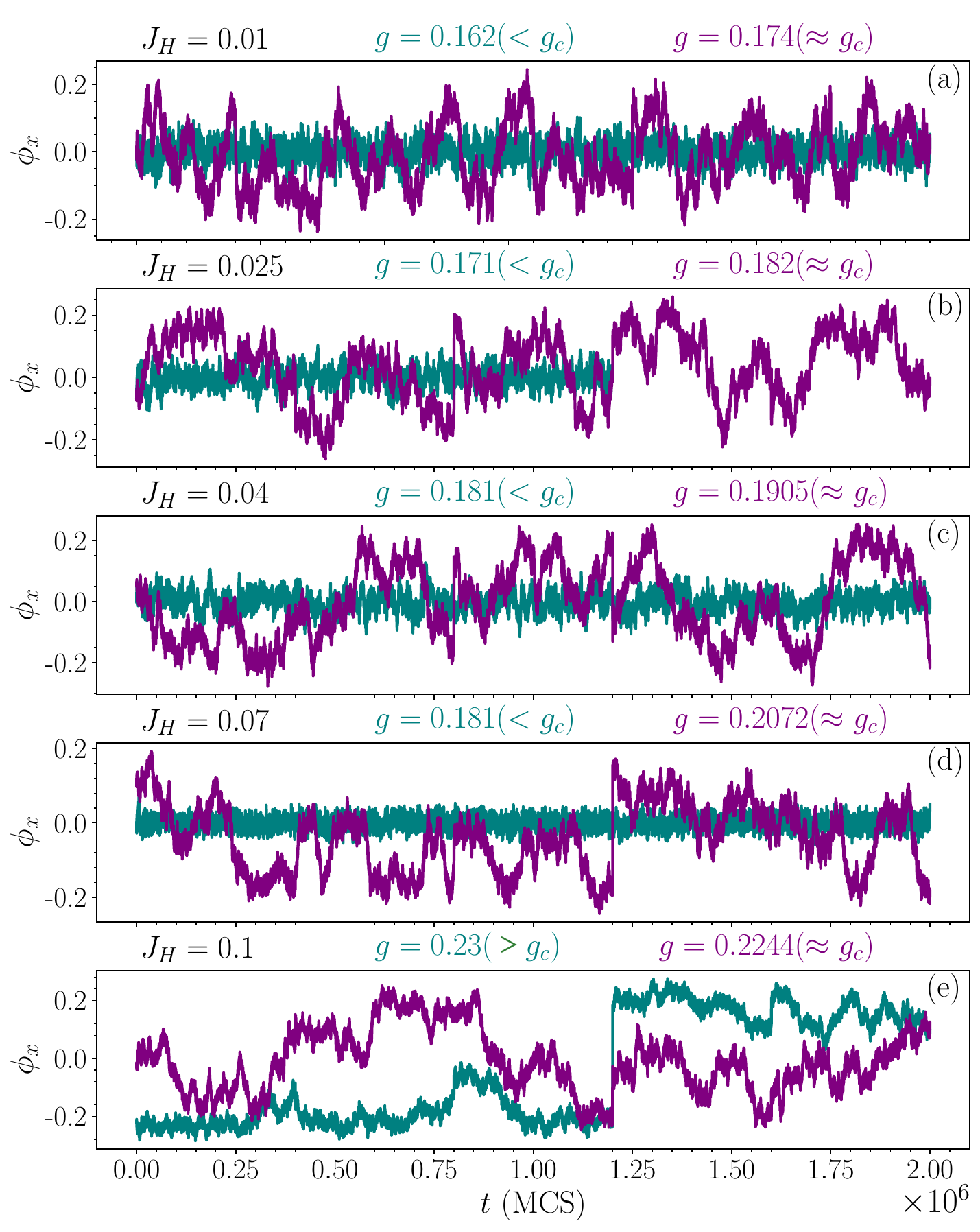}    
    \caption{ \label{timeseries_phix}
    This figure expands on Fig.~2 to document the various time series 
    data collected on cVBS order parameter.
     $\beta=\frac{L}{4}$ and $L=64$ throughout. The time series data
     are in register with those in Figs.~\ref{timeseries_phiy},\ref{timeseries_phi2}.
    }
\end{figure*}

\newpage
\begin{figure*}[h!]
    \centering

    \includegraphics[width=0.825\linewidth]{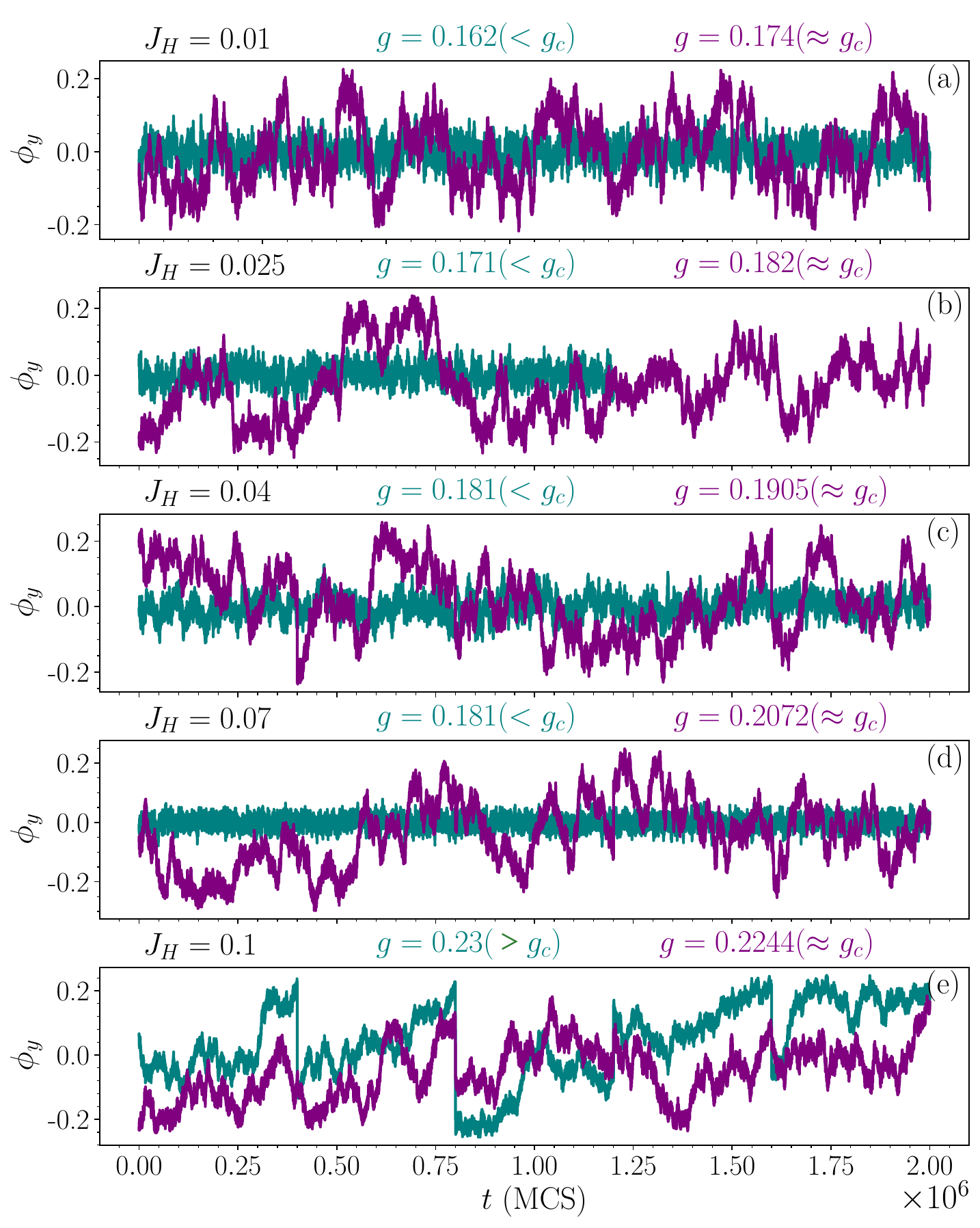}    
    \caption{ \label{timeseries_phiy}
    This figure expands on Fig.~2 to document the various time series 
    data collected on cVBS order parameter.
     $\beta=\frac{L}{4}$ and $L=64$ throughout.
     The time series data
     are in register with those in Figs.~\ref{timeseries_phix},\ref{timeseries_phi2}.
    }
\end{figure*}

\newpage
\begin{figure*}[h!]
    \centering

    \includegraphics[width=0.825\linewidth]{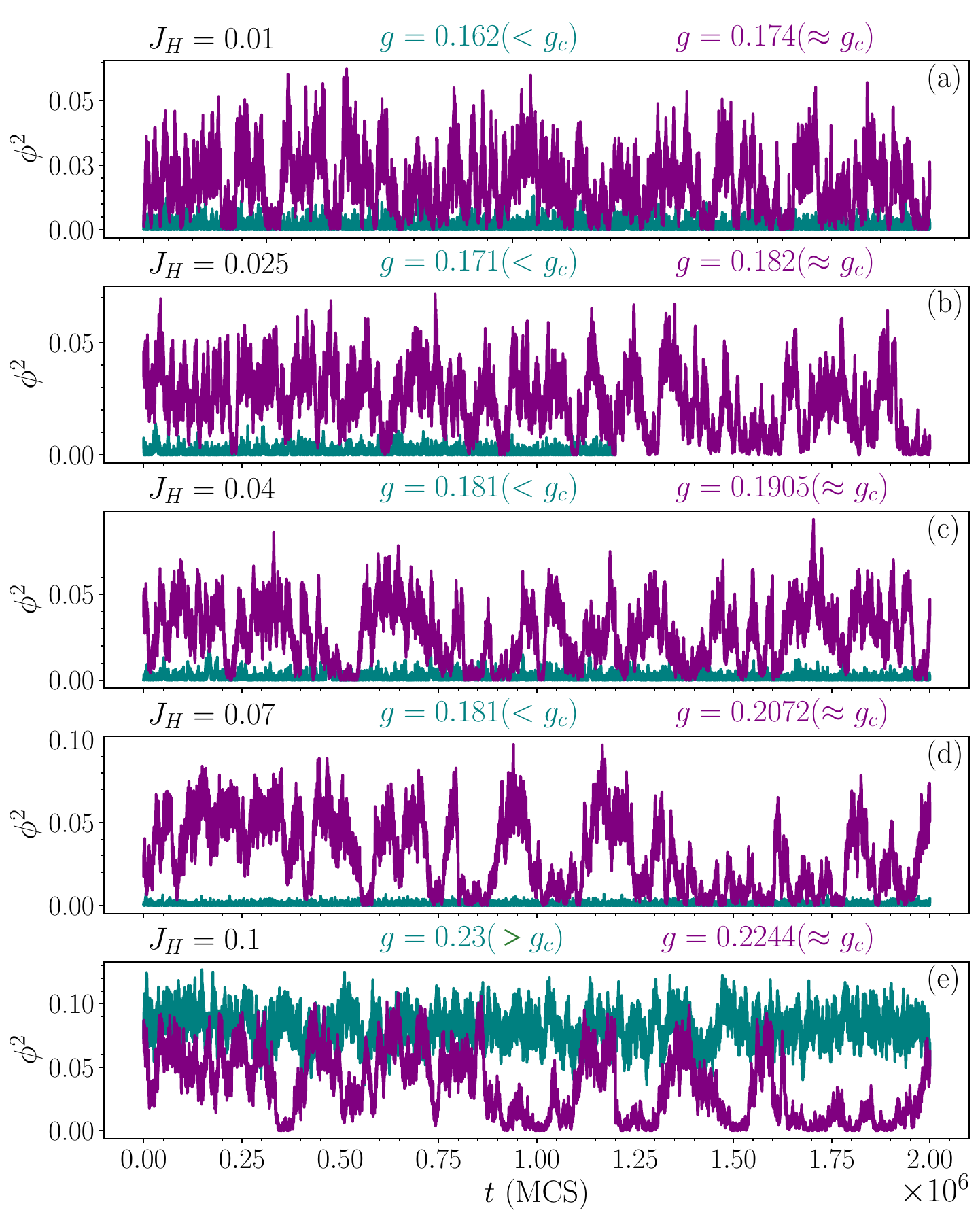}    
    \caption{\label{timeseries_phi2}
    This figure expands on Fig.~2 to document the various time series 
    data collected on cVBS order parameter.
     $\beta=\frac{L}{4}$ and $L=64$ throughout.
     The time series data
     are in register with those in Figs.~\ref{timeseries_phix},\ref{timeseries_phiy}.
    }
\end{figure*}

\clearpage
\pagebreak[4]
\subsection{Scaling collapse of correlation ratios and order parameters}
\label{subsec:collapse}

\begin{figure*}[h!]
    \centering
    \includegraphics[width=0.9\linewidth]{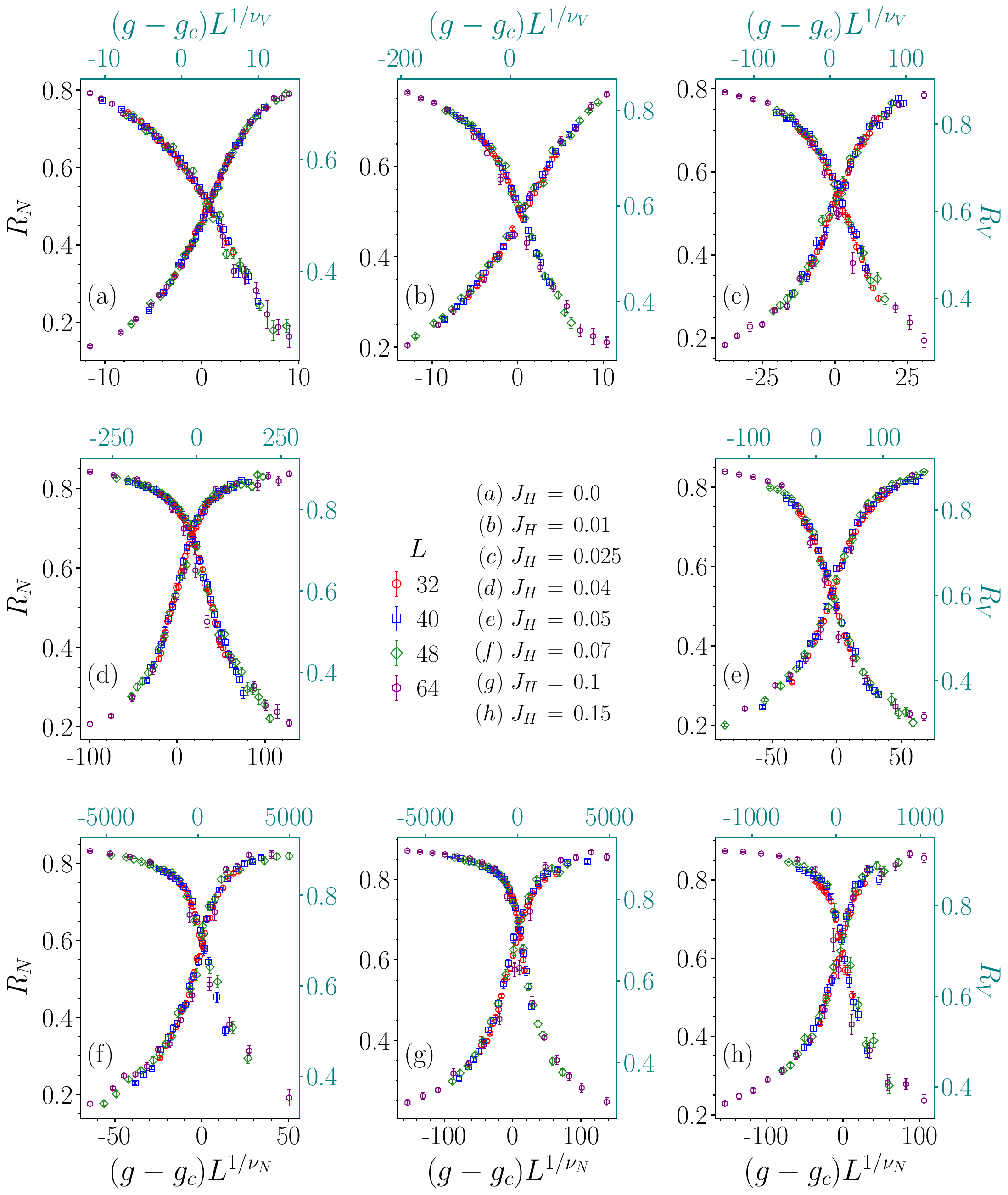} 
    \caption{ \label{order_param}
    This figure expands on Fig.~3a,b of the main text to document
    all the scaling collapse analysis performed on correlation ratios.  
    $\beta=\frac{L}{4}$ throughout.
    }
\end{figure*}

\begin{figure*}[h!]
    \centering
\includegraphics[width=0.9\linewidth]{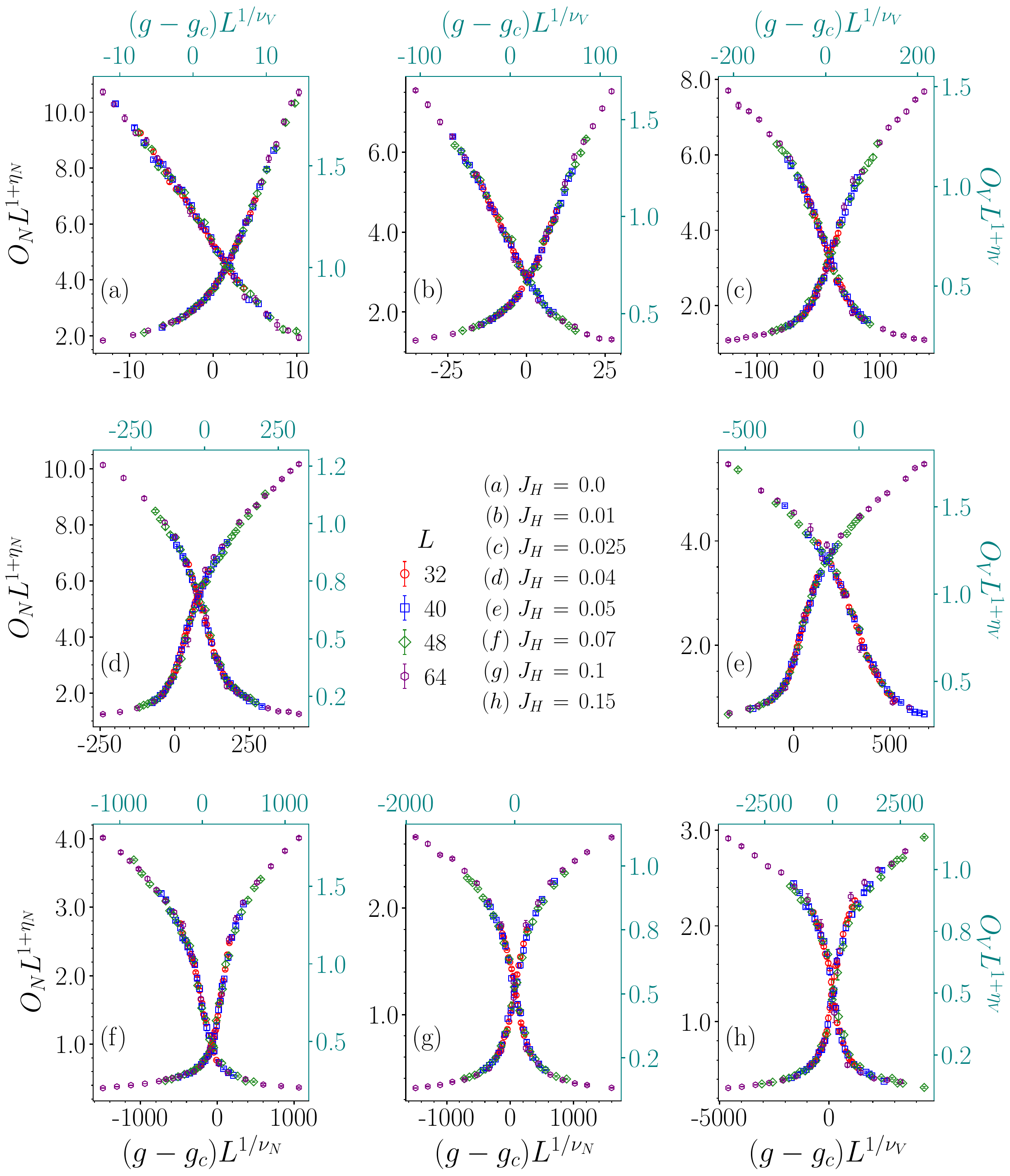}     
    \caption{ \label{order_param}
    This figure expands on Fig.~3c,d of the main text to document
    all the scaling collapse analysis performed on correlation ratios.  
    $\beta=\frac{L}{4}$ throughout. $\beta=\frac{L}{4}$ throughout.
    }
\end{figure*}

\clearpage
\pagebreak[4]
\subsection{Binder Ratios}
\label{subsec:corr_ratio}

\begin{figure*}[h!]
    \centering
    \includegraphics[width=0.9\linewidth]{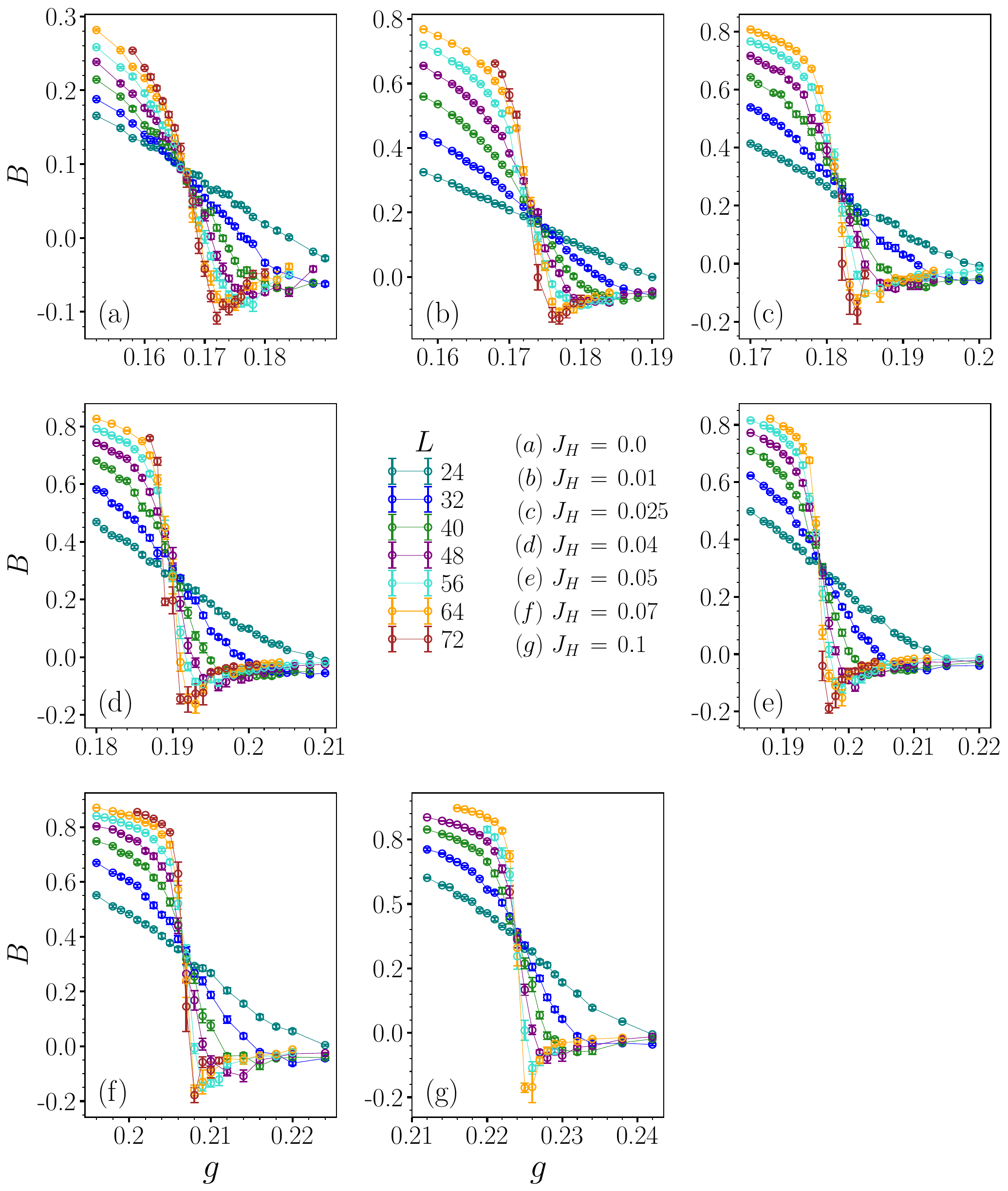}    
    \caption{\label{binder_plot}
    This figure expands on Fig.~4 of the main text to document all
    the Binder ratio data collected.
     $\beta=\frac{L}{4}$ throughout.
    }
\end{figure*}

\clearpage
\pagebreak[4]
\section{Tables}
\label{sec:exponent_tables}

\subsection{Critcal exponents and critical point estimates}

\begin{table*}[h]
\caption{Critical exponents ($\nu_N,\nu_V$) and critical point estimates
($g_{cN}$, $g_{cV}$) as obtained after performing a
scaling collapse on the N\'eel and cVBS correlation ratios ($R_{N}$ and $R_B$). 
System sizes used for collapses are in
the range of $L=24,32,40,48,64$. $\beta=\frac{L}{4}$ throughout.
Neither $(\nu_N,\nu_V)$ nor $(g_{cN},g_{cV})$ were set equal.
}
\begin{tabularx}{\textwidth}{*{9}{>{\centering\arraybackslash}X}}
\toprule
$J_H$ &  $\nu_{N}$ & $\nu_{V}$  & $g_{cN}$ & $g_{cV}$ & $\chi^2_N$ & $\chi^2_V$ \\
\toprule
0.0 &  0.49(5) & 0.63(1)  & 0.168(1) & 0.167(1) & 0.9-1.56 & 1.14-1.81\\
0.01 &   0.41(2) & 0.56(4)  & 0.175(1) & 0.171(1) & 1.31-2.27 & 1.32-1.84 \\
0.025 &  0.40(3) & 0.51(3)  & 0.182(1) & 0.18(1) & 0.97-2.43 & 1.15-1.61 \\
0.04 & 0.37(4) & 0.49(1)  & 0.191(1) & 0.188(1) & 1.12-2.64  & 1.14-1.86\\
0.05 &   0.40(3) & 0.46(3)  & 0.196(1) & 0.195(1) & 0.91-1.78 & 1.15-1.84 \\
0.07 &   0.34(5) & 0.48(5)  & 0.207(1) & 0.206(1)  & 1.52-2.54 & 1.04-1.77\\
0.1 &   0.31(1) & 0.40(1)  & 0.225(1) & 0.223(1) & 1.55-2.27 & 1.03-2.26\\
0.15 &   0.32(4) & 0.44(2)  & 0.254(1) & 0.252(1) & 1.37-3.39 & 0.74-1.91\\

\toprule
\end{tabularx}
\end{table*}
\begin{table*}[h]
\caption{Critical exponents ($\eta_V,\eta_N$, $\nu_N,\nu_V$) 
and critical point estimates ($g_{cN}$, $g_{cV}$)
as obtained after performing a scaling collapse of the 
N\'eel and cVBS order parameters ($O_{N},O_{B}$). 
System sizes used for collapses are in
the range of $L=24,32,40,48,64$.
$\beta=\frac{L}{4}$ throughout.
$(\nu_N,\nu_V)$ were not set equal to each other, while
$g_{cN}$ and $g_{cV}$ values were fixed while performing the
scaling collapse to the values obtained from
the scaling collapse of correlation ratios as shown the preceding table.}
\begin{tabularx}{\textwidth}{*{9}{>{\centering\arraybackslash}X}}
\toprule
$J_H$ &   $\nu_{N}$ & $\nu_{V}$  & $\eta_{N}$ & $\eta_{V}$ & $g_{cN}$ & $g_{cV}$ & $\chi^2_N$ & $\chi^2_V$ \\
\toprule
0.0 &   0.53(3) & 0.63(1) & 0.41(1) & 0.50(2) & 0.168 & 0.167 & 1.01-1.71 & 1.67-2.53\\
0.01 &  0.45(3) & 0.54(3) & 0.26(2) & 0.46(2) & 0.174 & 0.171 & 1.45-1.66 & 1.74-2.29\\
0.025 &   0.43(3) & 0.46(4) & 0.21(3) & 0.33(1)  & 0.182 & 0.180 & 1.04-1.55 & 0.86-1.37\\
0.04 &   0.40(2) & 0.43(5) & 0.24(2) & 0.23(2)  & 0.191 & 0.189 & 1.99-2.28 & 1.49-2.77\\
0.05 &   0.39(4) & 0.38(5) & 0.16(3) & 0.17(3)  & 0.196 & 0.195 & 1.28-1.54 & 0.98-1.9\\
0.07 &   0.38(2) & 0.36(2) & 0.05(1) & 0.12(3)  & 0.207 & 0.206 & 1.67-2.35 & 2.2-2.68\\
0.1 &   0.33(2) & 0.41(2) & 0.12(2) & 0.28(7)  & 0.225 & 0.223 & 2.48-2.81 & 2.55-2.86\\
0.15 &   ?? & ?? & ?? & ??  & 0.254 & 0.252 & ?? & ??\\

\toprule
\end{tabularx}
\end{table*}
\begin{table*}[h]
\caption{Critical exponents ($\eta_V,\eta_N$, $\nu_N,\nu_V$) 
and critical point estimates ($g_{cN}$, $g_{cV}$)
as obtained after performing a scaling collapse of 
N\'eel and cVBS order parameters ($O_{N},O_{B}$) 
for eight sets of the Heisenberg strength, $J_H=0.,0.01,0.025,0.04,0.05,0.07,0.1,0.15$. 
System sizes used for collapses are in
the range of $L=24,32,40,48,64$. $\beta=\frac{L}{4}$ throughout.
Neither $(\nu_N,\nu_V)$ nor $(g_{cN},g_{cV})$ were set equal.
This table was presented in the main text as well, and the
estimates of various fitting parameters below are corroborated well
by the estimates from the preceding tables.}
\begin{tabularx}{\textwidth}{*{9}{>{\centering\arraybackslash}X}}
\toprule
$J_H$ &  $\nu_{N}$ & $\nu_{V}$  & $\eta_{N}$ & $\eta_{V}$ & $g_{cN}$ & $g_{cV}$ & $\chi^2_N$ & $\chi^2_V$ \\
\toprule
0.0 &   0.53(3) & 0.63(1) & 0.44(5) & 0.49(2) & 0.168(1) & 0.167(1) & 1.08-1.68 & 1.69-2.46\\
0.01 &   0.45(2) & 0.54(3) & 0.23(3) & 0.42(4) & 0.174(1) & 0.171(1) & 1.19-1.63 & 1.38-1.73 \\
0.025 &   0.43(3) & 0.46(4) & 0.15(9) & 0.38(2)  & 0.182(1) & 0.180(1) & 0.75-1.46 & 0.8-1.4\\
0.04 &   0.40(2) & 0.43(5) & 0.13(7) & 0.30(8)  & 0.19(1) & 0.189(1) & 1.06-1.67 & 1.09-1.5\\
0.05 &   0.39(4) & 0.38(5) & 0.20(9) & 0.29(6)  & 0.196(1) & 0.195(1) & 0.87-1.31 & 0.87-1.96\\
0.07 &   0.38(2) & 0.39(3) & 0.10(4) & 0.10(4)  & 0.207(1) & 0.206(1) & 1.52-2.54 & 1.04-1.77\\
0.1 &   0.35(4) & 0.35(3) & -0.03(5) & -0.03(2)  & 0.224(1) & 0.224(1) & 1.24-3.28 & 0.99-1.97\\
0.15 &   0.33(2) & 0.33(1) & 0.00(8) & -0.12(8)  & 0.253(1) & 0.253(1) & 1.42-1.79 & 1.15-1.63\\

\toprule
\end{tabularx}
\end{table*}

\clearpage
\pagebreak[4]
\subsection{Benchmarking with Exact Diagonalization}
\begin{table*}[h]
\caption{This benchmarking table shows the values of total energy ($E$), 
N\'eel observables ($O_N$, $R_N$) obtained 
by Exact diagonalization (ED) and 
SSE-QMC on a $2\times2$ square plaquette at 
$\beta=10$.} 
\begin{tabularx}{\textwidth}{*{7}{>{\centering\arraybackslash}X}}
\toprule
$\left(J_{B},Q_B,J_{H}\right)$ 
& $E^{\text{ED}}$
& $E^{\text{SSE}}$ 
& $O^{\text{ED}}_N$ 
& $O^{\text{SSE}}_N$ 
& $R^{\text{ED}}_N$ 
& $R^{\text{SSE}}_N$  \\
\toprule
(1.0,1.0,0.2)   & -5.4524 & -5.452(5) & 1.6980  & 1.6980(2)  & 0.35739 & 0.3573(9)\\
(0.9,0.4,0.3)   & -3.6835 & -3.683(2) & 1.7386 & 1.738(5) & 0.36655 & 0.3665(4)\\
(0.6,0.5,0.5)   & -3.8492 & -3.849(0) & 1.7849 & 1.785(2) & 0.37651 & 0.3765(3)\\
(0.3,0.8,0.2)   & -3.4539 & -3.4539(6) & 1.7170 & 1.717(1) & 0.36172 & 0.3617(1)\\
(1.15,0.88,0.12)  & -5.2108 & -5.210(7) & 1.6861 & 1.686(2) & 0.35463 & 0.3546(0)\\
\toprule
\end{tabularx}
\end{table*}

\begin{table*}[h]
\caption{This benchmarking table shows the values of 
VBS observables 
($O_V$,  $R_V$, $O_B$, $R_B$) obtained by Exact diagonalization (ED) and SSE-QMC on a $2\times2$ square plaquette at $\beta=10$.}
\begin{tabularx}{\textwidth}{*{9}{>{\centering\arraybackslash}X}}
\toprule
$\left(J_{B},Q_B,J_{H}\right)$ 
& $O^{\text{ED}}_V$ 
& $O^{\text{SSE}}_V$ 
& $R^{\text{ED}}_V$
& $R^{\text{SSE}}_V$ 
& $O^{\text{ED}}_B$ 
& $O^{\text{SSE}}_B$ 
& $R^{\text{ED}}_B$ 
& $R^{\text{ED}}_B$ \\
\toprule
(1.0,1.0,0.2)  & 1.63654 & 1.6365(4) & 0.5 & 0.4999(5) & 1.49930 & 1.4993(3) & 0.5 & 0.5000(1)\\
(0.9,0.4,0.3)  & 1.65070 & 1.6507(4) & 0.5 & 0.4999(8) & 1.49615 & 1.496(2) & 0.5 & 0.49999(8)\\
 (0.6,0.5,0.5)  & 1.66564 & 1.665(5) & 0.5 & 0.49999(7) & 1.48881 & 1.488(8) & 0.5 & 0.4999(9)\\
(0.3,0.8,0.2)  & 1.64326 & 1.643(4) & 0.5 & 0.4999(9) & 1.49817 & 1.498(0) & 0.5 & 0.5000(3)\\
\scriptsize{(1.15,0.88,0.12)} & 1.63223 & 1.632(1) & 0.5 & 0.4999(5) & 1.49973 & 1.499(8) & 0.5 & 0.49999(3)\\
\toprule
\end{tabularx}
\end{table*}




